\documentclass[pdflatex,sn-basic]{sn-jnl}

\usepackage{acronym}
\usepackage{capt-of}
\usepackage[noabbrev,capitalise]{cleveref}
\usepackage[bottom]{footmisc}
\usepackage{mathtools}
\usepackage{multirow}
\usepackage{physics}
\usepackage{qcircuit-mod}
\usepackage[caption=false]{subfig}

\acrodef{ML}{machine learning}
\acrodef{QML}{quantum machine learning}
\acrodef{CNOT}{controlled-NOT}

\newcommand{\mathset}[1]{\mathcal{#1}}
\newcommand{\mathvec}[1]{\mathbf{#1}}

\jyear{2023}%

\theoremstyle{thmstyleone}%

\theoremstyle{thmstyletwo}%

\theoremstyle{thmstylethree}%

\begin{document}

\title[A quantum k-NN based on the Euclidean distance estimation]{A quantum k-nearest neighbors algorithm based on the Euclidean distance estimation}

\author*[1]{\fnm{Enrico} \sur{Zardini}}\email{enrico.zardini@unitn.it}

\author[1,2]{\fnm{Enrico} \sur{Blanzieri}}\email{enrico.blanzieri@unitn.it}

\author[2,3]{\fnm{Davide} \sur{Pastorello}}\email{davide.pastorello3@unibo.it}

\affil[1]{\orgdiv{Department of Information Engineering and Computer Science}, \orgname{University of Trento}, \orgaddress{\street{via Sommarive 9}, \city{Povo}, \postcode{38123}, \state{Trento}, \country{Italy}}}

\affil[2]{\orgname{Trento Institute for Fundamental Physics and Applications}, \orgaddress{\street{via Sommarive 14}, \city{Povo}, \postcode{38123}, \state{Trento}, \country{Italy}}}

\affil[3]{\orgdiv{Department of Mathematics}, \orgname{Alma Mater Studiorum - Università di Bologna}, \orgaddress{\street{Piazza di Porta San Donato 5}, \city{Bologna}, \postcode{40126}, \country{Italy}}}

\abstract{The $k$-nearest neighbors ($k$-NN) is a basic machine learning (ML) algorithm, and several quantum versions of it, employing different distance metrics, have been presented in the last few years. Although the Euclidean distance is one of the most widely used distance metrics in ML, it has not received much consideration in the development of these quantum variants. In this article, a novel quantum $k$-NN algorithm based on the Euclidean distance is introduced. Specifically, the algorithm is characterised by a quantum encoding requiring a low number of qubits and a simple quantum circuit not involving oracles, aspects that favor its realization. In addition to the mathematical formulation and some complexity observations, a detailed empirical evaluation with simulations is presented. In particular, the results have shown the correctness of the formulation, a drop in the performance of the algorithm when the number of measurements is limited, the competitiveness with respect to some classical baseline methods in the ideal case, and the possibility of improving the performance by increasing the number of measurements.}

\keywords{quantum computing, quantum machine learning, k-nearest neighbors, Euclidean distance}

\maketitle

\section{Introduction}
\label{sec:introduction}
\Ac{QML} is one of the most recent and most popular directions of scientific investigation in the area of quantum computing. In particular, the application of quantum computation to \ac{ML} tasks offers some interesting solutions characterised by a quantum advantage with respect to the classical counterparts, at least on a theoretical level. Furthermore, \ac{QML} seems to be a good way to exploit existing prototypes of quantum computers for tackling real-world problems. In this sense, a general ``practical" approach consists in executing quantum algorithms as subroutines of more complex learning schemes, in which a quantum machine is used as a co-processor within a hybrid architecture. This approach is an interesting alternative to the development of quantum algorithms that fully accomplish \ac{ML} tasks under the (strong) assumptions of ideality and universality of the quantum hardware.

In the last decade, several interesting \ac{QML} algorithms have been proposed and characterised from a theoretical viewpoint; sometimes, they have been also empirically tested. Remarkable examples are the quantum SVM proposed by \cite{SVMLloyd}, the distance-based classifiers like the one defined by \cite{Schuld_2017}, and the quantum neural networks, whose performance have been discussed by \cite{Abbas_2020}. In particular, several quantum versions of the $k$-nearest neighbors ($k$-NN) algorithm have been proposed (see section \ref{sec:background}). In \ac{ML}, the $k$-NN is a very simple and widely used classification algorithm that assigns a label to an unclassified data instance according to the labels of the $k$ nearest training instances. To do so, a suitable reference distance in the space in which the data are represented must be selected. In the classical realm, typical choices are the Hamming distance and the Euclidean distance; instead, in the quantum realm (considering only the quantum $k$-NNs), the Euclidean distance has not received much consideration. In this work, we propose a quantum $k$-nearest neighbors algorithm in which the calculation of the Euclidean distances is based on a novel quantum encoding with low qubit requirements and a simple quantum circuit, making the implementation particularly advantageous. As with other algorithms involving the quantum computation of the Euclidean distance (e.g., by means of the SWAP test), an exponential speedup over a classical calculation can be obtained assuming the availability of a quantum random access memory \citep[QRAM,][]{qram} for data retrieval. Otherwise, there is not a true quantum advantage in terms of time complexity. From a practical viewpoint, in this article, we analyse the performance of the proposed quantum $k$-NN in terms of classification accuracy and correctness of the nearest neighbors found (evaluated through the Jaccard index). The algorithm has been implemented using Qiskit and run with three different execution modalities: \textit{classical}, \textit{statevector}, and \textit{simulation}. Instead, the empirical evaluation on a real quantum machine was prevented by the number of qubits required by the considered experiments.

The article is structured as follows: \cref{sec:background} provides some background information; \cref{sec:method} presents the new quantum $k$-nearest neighbors algorithm based on the Euclidean distance metric; \cref{sec:implementation} deals with the implementation of the algorithm; \cref{sec:empirical-evaluation} describes the experimental evaluation and the results obtained; \cref{sec:conclusion} concludes the article.

\section{Background}
\label{sec:background}
This section presents background information about quantum machine learning,
the quantum $k$-nearest neighbors algorithms available in the literature, and the usages of the (squared) Euclidean distance in the field of quantum machine learning.

\subsection{Quantum machine learning}
\label{subsec:qml}
In general, \ac{ML} is the automation of methodologies for extracting information from collected data. If the data analysis techniques are implemented on conventional digital computers, we refer to classical \ac{ML}; if quantum machines are employed, we refer to quantum \ac{ML}. A general reason justifying the efforts in developing new \ac{QML} schemes is suggested by \citet{biamonte}: since (even small) quantum systems are difficult to simulate with classical computers, we can conjecture that (even small) quantum processors can find structures in data that are difficult to discover classically. Therefore, \ac{QML} could be the right path towards non-trivial applications of the small-scaled quantum machines available today and in the near future. On the other hand, under strong assumptions of universality, large scale, and fault tolerance, it is possible to formulate several \ac{QML} algorithms that outperform their classical counterparts. This is very important for the comprehension of the foundations of quantum computing and for showing the actual potential of quantum computers. However, to promote the advent of quantum technologies in the near term, taking into account the limitations of the current quantum hardware is useful while seeking new \ac{QML} schemes. 

From the mathematical viewpoint, there is another relevant motivation for developing \ac{ML} algorithms to be executed by quantum machines, given by a formal analogy between quantum mechanics and \ac{ML}: both fields rely on matrix operations in high-dimensional vector spaces. In practice, the Hilbert spaces, in which physical quantum systems are described, can be used as feature spaces for data representations. In this framework, linear algebraic operations are physically realized by the time evolution of quantum states; for instance, in the circuit model of quantum computation, the evolution is described as the action of quantum gates, i.e., unitary operators. In addition, representing data into quantum states is advantageous also in terms of space resources, since the dimension of the Hilbert space of a multi-qubit system is exponential in the number of qubits. Then, the controlled dynamics of a small number of qubits towards a target state may correspond to the application of a complex linear algebraic operation on the considered feature space.

A crucial notion in \ac{QML} is \textit{quantum encoding}, which is any procedure that encodes classical data (e.g., a list of symbols) into quantum states. In particular, loading efficiently large amounts of data into quantum architectures is a serious bottleneck at the current status of QML; indeed, the state preparation required for running several well-known QML algorithms can be done efficiently only under the strong assumption of the availability of a QRAM. More in detail, given an $n$-qubit register, let $\{\ket{i}\}_{i=0,...,2^n-1}$ be a fixed orthonormal basis of the corresponding Hilbert space that we call \textit{computational basis}. The simplest quantum encoding is the \textit{basis encoding}, in which the bit strings of length $n$ are encoded into the states that form the computational basis. Therefore, $n$ qubits are used to encode $n$ bits of classical information with interesting quantum opportunities, like creating superpositions of data and enabling non-classical correlations via entanglement. Instead, a more efficient quantum encoding in terms of space resources is the \textit{amplitude encoding}, in which a data instance represented by a normalized complex vector $\mathbf{x} \in \mathbb{C}^{2^n}$ is encoded into the coordinates (or amplitudes) of a quantum state with respect to the computational basis, namely,
\begin{equation*}
    \ket\psi=\sum_{i=0}^{2^n-1}x_i \ket i\qquad (n\mbox{-qubit state}).
    \label{eq:amplitude-encoding}
\end{equation*}
The amplitude encoding exploits the exponential storing capacity of a quantum memory, but it does not allow the direct retrieval of the stored data. Indeed, the amplitudes cannot be observed, and only the probabilities $\lvert x_i\rvert^2$ can be estimated. The encoding procedure used in this work is based on amplitude encoding. 

Let us conclude this introductory section by arguing that \ac{QML} is probably the most promising way to find out effective applications of the existing small-scale quantum computers. In particular, one can also drop the requirement that an \ac{ML} task must be entirely accomplished by quantum computations in favor of hybrid approaches, in which quantum co-processors efficiently solve specific subproblems within more complex learning schemes. Moreover, the quantum speedup is not the unique quantum advantage that can be pursued. Accuracy in prediction, expressive power, generalisation capability, and the ability to avoid plateaus in training are also noteworthy figures of merit in evaluating the learning performance of quantum machines.

\subsection{Quantum k-NN}
\label{subsec:quantum-k-nn}
The $k$-nearest neighbors \citep{k_nn} is a classification algorithm that consists of three steps: the computation of the distance with respect to the training elements; the identification of the $k$ nearest neighbors, i.e., the $k$ elements closest to the test instance; the prediction of the class label through a majority voting. Several quantum variants with different distance measures have been proposed, but a common aspect to all of them is the exploitation of a superposition state in order to perform parallel operations, such as computing the distance from the training elements simultaneously (quantum parallelism).

First of all, quantum $k$-NN algorithms employing the Hamming distance, thus requiring binary features, have been proposed by \citet{schuld_qknn_pattern_class}, \citet{qknn_herm_matrix_rec}, \citet{ruan2017_qknn_hamming}, \citet{qknn_img_class_kl_transf}, and \citet{li_2021_qknn}. Specifically, the first two works compute the Hamming distances by encoding the sums of the qubits differences (differences obtained through controlled-NOT gates) into the amplitudes by means of a unitary operation \citep[an idea proposed first by][]{trugenberger_q_sum}. Then, the classification is performed directly by measuring without explicitly selecting the nearest neighbors. Instead, the other works exploit the incrementation circuit presented by \citet{incrementation_circuit} in order to obtain the distance values in basis encoding. After that, \citet{ruan2017_qknn_hamming} select the data with a distance lower than a given threshold by means of an OR gate and a projection operation to directly perform the classification, \citet{qknn_img_class_kl_transf} exploit D\"{u}rr's minimization algorithm \citep{durr_minimization} to find the $k$ minimum distance values, while \citet{li_2021_qknn} apply a novel quantum search procedure inspired by a binary search in order to identify the minimum.

Concerning non-binary features, distance measures related to the angle between vectors, such as the cosine distance, are widely used. For instance, \citet{qknn_img_class} and \citet{qknn_img_digit_rec} have applied a quantum $k$-NN variant of this type to image classification tasks. In particular, the SWAP test \citep{swap_test} without measurements is used to compute the distances, whose values are then transferred to the qubits states through the amplitude estimation algorithm \citep{ampl_ampl_and_estim}. Finally, the nearest neighbors are found by means of D\"{u}rr's algorithm. This workflow has been presented first by \citet{q_nearest_neighbor}, although for finding only the nearest neighbor. Instead, \citet{qknn_implemented} and \citet{qknn_cat_tensor_networks} have proposed a conceptually simpler variant, which consists in iterating SWAP tests and measurements in order to estimate a quantity proportional to the squared cosine similarity with respect to the training instances. In addition, the model allows processing multiple test instances in parallel, as shown by \citet{qknn_cat_tensor_networks}. Actually, \citeauthor{qknn_implemented} have recently proposed another variant \citep{basheer2021quantum} whose workflow, however, is not so different from that of the previously described works. Indeed, it involves the SWAP test, a quantum analog-to-digital conversion algorithm \citep{q_analog_to_dig_conv}, and a variation of D\"{u}rr's algorithm.

Other interesting distance measures for non-binary features are the Euclidean, the Mahalanobis, and the polar distances. Specifically, the Euclidean distance is dealt with in depth in \cref{subsec:quantum-euclidean-dist}. Regarding the other ones, \citet{Gao2022_Mahalanobis} have proposed a variant based on the Mahalanobis distance, while \citet{Feng_2023_Polar} have presented a quantum $k$-NN based on the polar distance, which combines angle and module length information through an adjustable parameter. In detail, the Mahalanobis distance is computed by exploiting the phase estimation algorithm \citep{phase_estimation}, combined with Hamiltonian simulation \citep{hamiltonian_simulation}, and a controlled rotation; instead, the polar distance is calculated through the SWAP test without measurements and a pair of Toffoli gates (one of which extended). After that, in both works, the distances are encoded in the qubits states by applying the amplitude estimation algorithm \citep[or its coherent version, proposed by][]{q_nearest_neighbor} and the nearest neighbors are retrieved through D\"{u}rr's algorithm \citep[or an algorithm based on it, proposed by][]{Miyamoto_minimum_finding}. To conclude, it is also worth mentioning the quantum $k$-NN, based on a quantum sorting subroutine, that has been proposed by \citet{Quezada_2022_Sorting_Qknn}. It requires a metric operator computing distances and encoding them in qubits states, an oracle that identifies sorted sequences, and Grover's algorithm \citep{grover}; as in other works, the classification is performed directly without identifying the $k$ nearest neighbors.

\subsection{Quantum Euclidean distance}
\label{subsec:quantum-euclidean-dist}
The Euclidean distance is a well-known distance metric in \ac{ML}. Here, the definition of its squared version is provided, since it will be useful in the following sections. In particular, given two vectors  $\mathvec{u}, \mathvec{v} \in \mathbb{R}^n$, the squared Euclidean distance between them, $d^2(\mathvec{u},\mathvec{v})$, is defined as
\begin{equation}
    d^2(\mathvec{u},\mathvec{v}) = \| \mathvec{u} - \mathvec{v} \|^2 = \|\mathvec{u}\|^2 - 2 \langle \mathvec{u}, \mathvec{v}\rangle + \|\mathvec{v}\|^2,
    \label{eq:sq-eucl-dist}
\end{equation}
where $\langle \mathvec{u}, \mathvec{v}\rangle$ is the scalar product between $\mathvec{u}$ and $\mathvec{v}$.

The distance metric in question has been employed also in the field of \ac{QML}. For instance, \citet{Lloyd_2013} have proposed a quantum procedure to estimate the squared Euclidean distance between a data point and the centroid of a cluster, i.e., the mean of the elements contained in a group of data. Specifically, the algorithm relies on the SWAP test, which is applied to the index registers, and does not require the input vectors to have unit norms. An analogous procedure has been used by \citet{Sarma_2019} to provide a hybrid $k$–means clustering algorithm (in which the centroids are classically computed), and by \citet{Getachew_2020} for a hybrid version of the $k$-medians one. Instead, \cite{Yu2020} have proposed three quantum algorithms to estimate three similarity measurements, based on the squared Euclidean distance, for data sets. In particular, all procedures do not require unit-norm input vectors, exploit the quantum interference (given by the change of basis), and use the amplitude estimation algorithm to determine the similarity measures. Finally, it is worth mentioning the quantum binary classifier devised by \cite{Schuld_2017}. In detail, the classifier circuit consists of a Hadamard gate (necessary for the quantum interference), a conditional measurement, and a final measurement. By iterating the procedure just described, a probability value related to the squared Euclidean distances is estimated for each class. In this last work, input vectors with unit norms are considered. 

Regarding the quantum $k$-NN model, as far as the authors know, the only variant based on the Euclidean distance available in the literature has been presented by \citet{basic_qknn}; it exploits the procedure proposed by \citet{Lloyd_2013} to estimate the pairwise distance values, and D\"{u}rr's minimization algorithm to find the $k$ nearest neighbors. The main drawback lies in the need of multiple iterations for each of these steps, since both involve a final measurement. In addition, D\"{u}rr's algorithm requires an oracle, i.e., a black-box function, to be used. Actually, the nearest neighbor algorithm proposed by \cite{q_nearest_neighbor} admits also the Euclidean distance as the distance metric. However, the workflow, which has been described in \cref{subsec:quantum-k-nn}, is quite complex to be implemented. Finally, the computation of the single linkage, namely, one of the set similarity measures considered by \cite{Yu2020}, could be seen as a generalisation of the nearest neighbor search. Nevertheless, their quantum algorithm uses the reciprocals of the input vectors; as a consequence, theoretically, the original distance relationships are not preserved.

\section{Method}
\label{sec:method}
In this section, the new quantum $k$-NN algorithm based on the Euclidean distance metric is presented. In addition, a brief discussion of the algorithm's complexity compared with that of the classical counterpart is provided.

\subsection{Algorithm}
\label{subsec:algorithm}
In the quantum $k$-NN algorithm introduced in this work, a quantity related to the squared Euclidean distance is computed in parallel for all training instances by means of a novel encoding and a simple quantum circuit, which performs a SWAP-test-like procedure without controlled-SWAP gates. In practice, the algorithm exploits the quantum interference and encodes these distance-related values, which are then estimated through measurements, in the amplitudes of the quantum states. It is worth highlighting that the input vectors do not undergo a unit-norm normalization, which would result in a significant information loss. In addition, the number of qubits needed is low, and no oracle is involved, making the implementation feasible. A more detailed and formal description of the steps of this new algorithm is provided below.

\subsubsection{Data preprocessing}
\label{subsubsec:data-preprocessing}
Let us consider a training set $\mathset{U} = \{\mathvec{u}_0, ..., \mathvec{u}_{N-1} \}$ of real-valued data instances $\mathvec{u}_j \in \mathbb R^d$, and let $\mathset{L} = \{l_0, ..., l_{N-1} \}$ be the set of corresponding labels. In addition, let us consider a test instance $\mathvec{u}' \in \mathbb R^d$, whose label is unknown.

The preprocessing step of the algorithm consists in normalizing the data features into the range $\left[-\frac{1}{2\sqrt{d}}, \frac{1}{2\sqrt{d}}\right]$. In this way, the maximum norm of the resulting vectors turns out to be $\frac{1}{2}$ and the maximum (squared) Euclidean distance turns out to be 1.

\subsubsection{Initial state and encoding(s)}
\label{subsubsec:initial-state-and-encodings}
Let $\mathset{V} = \{\mathvec{v}_0, ..., \mathvec{v}_{N-1} \}$ and $\mathvec{v}'$ be the training set and the test instance after the preprocessing step described above. The quantum circuit is then initialized in the state
\begin{equation}
    \ket{\psi} = \ket{0} \otimes \left(\frac{1}{\sqrt{2}} (\ket{0}\ket{\alpha} + \ket{1}\ket{\beta})\right),
    \label{eq:initial-psi-state}
\end{equation}
where
\begin{equation*}
    \ket{\alpha} = \frac{1}{\sqrt{N}} \sum_{j=0}^{N-1}\ket{j} \sum_{i=0}^{F-1}x_{ji}\ket{i},
    \label{eq:alpha-state}
\end{equation*}
\begin{equation*}
    \ket{\beta} = \frac{1}{\sqrt{N}} \sum_{j=0}^{N-1}\ket{j} \sum_{i=0}^{F-1}x'_{ji}\ket{i}.
    \label{eq:beta-state}
\end{equation*}
Here, $F$ is a positive integer value depending on the encoding used, while $\mathvec{x}_j = \{ x_{ji} \}_{i=0,...,F-1}$ and $\mathvec{x}'_j = \{ x'_{ji} \}_{i=0,...,F-1}$ represent the quantum encoded versions of the preprocessed training and test data, respectively. Therefore, the number of qubits required is $2 + \lceil \log_2 N \rceil + \lceil \log_2 F \rceil$. In particular, two encodings, whose advantages are discussed in the next sections, have been devised and tested in this work: \textit{extension} and \textit{translation}. Let us look at their definitions. As regards the \textit{extension} encoding, $F = 2d + 3$ and

\noindent\begin{minipage}{0.35\linewidth}
    \begin{equation*}
        x_{ji}=\left\{
            \def\arraystretch{1.5}\begin{array}{l}       
                \frac{2}{\sqrt{3}}v_{ji}                 \\   
                \frac{2}{\sqrt{3}}v_{j(i-d)}             \\   
                \frac{2}{\sqrt{3}}\|\mathvec{v}_j\|      \\   
                0                                        \\   
                \sqrt{1 - 4\|\mathvec{v}_j\|^2}               
            \end{array}\right.
        \label{eq:extension-x}
    \end{equation*}
\end{minipage}%
\begin{minipage}{0.60\linewidth}
    \begin{equation*}
        x'_{ji}=\left\{
            \arraycolsep=9pt\def\arraystretch{1.5}\begin{array}{ll}
                -\frac{2}{\sqrt{3}}v'_{i}                                          & 0 \le i < d    \\  
                -\frac{2}{\sqrt{3}}v'_{(i-d)}                                      & d \le i < 2d   \\
                \frac{2}{\sqrt{3}}\|\mathvec{v}_{j}\|                              & i = 2d         \\
                \sqrt{1 - \frac{4}{3}(2\|\mathvec{v}'\|^2 + \|\mathvec{v}_j\|^2)}  & i = 2d + 1     \\
                0                                                                  & i = 2d + 2,          
            \end{array}\right.
        \label{eq:extension-x'}
    \end{equation*}
\end{minipage}

\vspace{15pt}\noindent with $v_{ji}$ being the $i$-th feature of the $j$-th preprocessed training instance, and $v'_i$ being the $i$-th feature of the preprocessed test instance. Instead, for the \textit{translation} encoding, $F = 2d + 4$ and

\noindent\begin{minipage}{0.35\linewidth}
    \begin{equation*}
        x_{ji}=\left\{
            \def\arraystretch{1.5}\begin{array}{l}              
                v_{ji}                                          \\   
                v_{j(i-d)}                                      \\   
                \|\mathvec{v}_j\|                               \\   
                \frac{1}{2}                                     \\   
                0                                               \\   
                \sqrt{\frac{3}{4} - 3\|\mathvec{v}_j\|^2}             
            \end{array}\right.
        \label{eq:translation-x}
    \end{equation*}
\end{minipage}
\begin{minipage}{0.60\linewidth}
    \begin{equation*}
        x'_{ji}=\left\{
            \arraycolsep=9pt\def\arraystretch{1.5}\begin{array}{ll}
                -v'_{i}                                                            & 0 \le i < d    \\  
                -v'_{(i-d)}                                                        & d \le i < 2d   \\
                \|\mathvec{v}_j\|                                                  & i = 2d         \\
                -\frac{1}{2}                                                       & i = 2d + 1     \\
                \sqrt{\frac{3}{4} - (2\|\mathvec{v}'\|^2 + \|\mathvec{v}_j\|^2)}   & i = 2d + 2     \\
                0                                                                  & i = 2d + 3.          
            \end{array}\right.
        \label{eq:translation-x'}
    \end{equation*}
\end{minipage}

\vspace{15pt}\noindent As a consequence, the number of qubits required is the same for both encodings. It is also worth highlighting that, in both cases, $\mathvec{x}_j$ (and therefore $\ket{\alpha}$) is independent of the preprocessed test instance $\mathvec{v}'$, whereas $\mathvec{x}'_j$ (and therefore $\ket{\beta}$) depends on the preprocessed training set $\mathset{V}$. 

\subsubsection{Bell-H operation and final state}
\label{subsubsec:bell-h-operation-and-final-state}
After the initial state preparation, an operation denoted here as \textit{Bell-H} is performed. In detail, the \textit{Bell-H} corresponds to a SWAP-test-like procedure in which the states of interest ($\ket{\alpha}$ and $\ket{\beta}$), initially prepared in superposition, interfere by means of a \ac{CNOT} gate. The corresponding quantum circuit, including also the initial state preparation (separated by a dashed vertical line), is the following:
\vspace{5pt}

$$
    \Qcircuit @C=1em @R=0.5em @!R {
        \lstick{\ket{0}}             & \qw                                                     & \qw \ar@{--}[]+<0em,0.5em>;[dd]+<0em,-0.5em> \raisebox{3em}{$\ket{\psi}$}               
                                     & \gate{H} & \ctrl{1} & \gate{H} & \qw & & \\ 
        \lstick{\ket{0}}             & \multigate{1}{\parbox{1.5cm}{\centering State\\ init.}} & \qw 
                                     & \qw      & \targ    & \qw      & \qw & & \\  
        \lstick{\ket{0}^{\otimes I}} & \ghost{\parbox{1.5cm}{\centering State\\ init.}}        & \qw    
                                     & \qw      & \qw      & \qw      & \qw & & \rstick{\raisebox{3.85em}{$\ket{\gamma},$}} \gategroup{1}{8}{3}{8}{.4em}{\}}
    }    
$$

\vspace{8pt}\noindent where $I = \lceil \log_2 N \rceil + \lceil \log_2 F \rceil$. In practice, the \textit{Bell-H} circuit consists of a Hadamard gate applied to the first qubit, a \ac{CNOT} gate with the first qubit as control and the second qubit as target, and another Hadamard gate applied to the first qubit. Hence, the difference with respect to a standard Bell circuit, commonly used to generate Bell states, lies in the presence of an additional downstream Hadamard gate. A significant advantage with respect to the standard SWAP test lies in the constant number of elementary gates required (three), independently of the size of the states involved; as a drawback, the preparation of the input state is more complex, especially without the availability of a QRAM. 

The output state obtained after the \textit{Bell-H} operation is
\begin{multline}
    \ket{\gamma} = \frac{1}{2} \left(\ket{0} \otimes \left(\frac{1}{\sqrt{2}}(\ket{0}\ket{\alpha} + \ket{0}\ket{\beta} + \ket{1}\ket{\beta} + \ket{1}\ket{\alpha})\right)\right. + \\
    \left.\ket{1} \otimes \left(\frac{1}{\sqrt{2}}(\ket{0}\ket{\alpha} - \ket{0}\ket{\beta} + \ket{1}\ket{\beta} - \ket{1}\ket{\alpha})\right)\right),
    \label{eq:output-gamma-state}
\end{multline}
and the probability of measuring $1$ on the first qubit is $\frac{1}{2}(1 - \bra{\alpha}\ket{\beta})$ (the derivation is shown in \cref{subsec:app-state-one-prob}). By pulling out the summation on the index register ($\ket{j}$ inside $\ket{\alpha}$ and $\ket{\beta}$) and tracing out (i.e., discarding) the second qubit and the features register ($\ket{i}$ inside $\ket{\alpha}$ and $\ket{\beta}$), it is possible to write the final state as
\begin{equation}
    \ket{\delta} = \frac{1}{\sqrt{N}} \sum_{j=0}^{N-1}\left[\sqrt{1 - s(\mathvec{v}_j, \mathvec{v}')}\ket{0} + \sqrt{s(\mathvec{v}_j, \mathvec{v}')}\ket{1}\right]\ket{j},
    \label{eq:final-delta-state}
\end{equation}
with $s(\mathvec{v}_j, \mathvec{v}')$ being a similarity measure related to the squared Euclidean distance between $\mathvec{v}_j$ and $\mathvec{v}'$; hence, the lower the distance, the higher the $s(\mathvec{v}_j, \mathvec{v}')$ value. Specifically, $s(\mathvec{v}_j, \mathvec{v}')$ is given by
\begin{equation}
    s(\mathvec{v}_j, \mathvec{v}') = P(\mathit{qubit}_1 = 1 \mid j) = \frac{1}{2}(1 - \langle \mathvec{x}_j, \mathvec{x}'_j\rangle),
    \label{eq:s-value}
\end{equation}
where $\mathit{qubit}_1$ is the first qubit in the circuit, and the value of $\langle \mathvec{x}_j, \mathvec{x}'_j\rangle$ depends on the encoding used, as shown in \cref{tab:encodings-properties} (a more detailed description of how Equation \ref{eq:final-delta-state} is obtained is provided in \cref{subsec:app-reduced-final-state}). 

\begin{table}[htb]
    \centering
    \caption{Properties of the two encodings. Notice that the range of values of $s(\mathvec{v}_j, \mathvec{v}')$ is determined by the preprocessed test instance $\mathvec{v}'$}
    \label{tab:encodings-properties}
    \begin{tabular}{c|c|c}
                                                              & Extension                                            & Translation                                           \rule[-5pt]{0pt}{15pt} \\ \hline
        $\langle \mathvec{x}_j, \mathvec{x}'_j\rangle$ value  & $\frac{4}{3}(\|\mathvec{v}_j\|^2 - 2 \langle \mathvec{v}_j, \mathvec{v}'\rangle)$  & $\|\mathvec{v}_j\|^2 - 2 \langle \mathvec{v}_j, \mathvec{v}'\rangle - \frac{1}{4}$  \rule[-5pt]{0pt}{15pt} \\ \hline
        Minimum $s(\mathvec{v}_j, \mathvec{v}')$ range        & [0.333, 0.5]                                         & [0.5, 0.625]                                          \rule[-5pt]{0pt}{15pt} \\ \hline
        Maximum $s(\mathvec{v}_j, \mathvec{v}')$ range        & [0, 0.666]                                           & [0.25, 0.75]                                          \rule[-5pt]{0pt}{15pt} \\
    \end{tabular}
\end{table}

Looking at the first row of \cref{tab:encodings-properties} and recalling \cref{eq:sq-eucl-dist}, it is possible to notice two aspects: $\langle \mathvec{x}_j, \mathvec{x}'_j\rangle$ is strictly related to the squared Euclidean distance between $\mathvec{v}_j$ and $\mathvec{v}'$ for both encodings; the term $\|\mathvec{v}'\|^2$ does not appear. However, the latter is not an issue, since $\|\mathvec{v}'\|^2$ is the same for all training instances. The other information contained in the table allow understanding the strong point of each encoding. In particular, the \textit{extension} encoding maximises the range of possible values of $s(\mathvec{v}_j, \mathvec{v}')$, allowing a better representation of the similarity values, namely, a representation less sensitive to the presence of noise. Instead, the \textit{translation} encoding maximises the probability of measuring 1 on the first qubit, a favorable situation for reasons that will become clear in the next section. Eventually, it is worth highlighting that the range of $s(\mathvec{v}_j, \mathvec{v}')$ is determined by $\mathvec{v}'$; in detail, the minimum range corresponds to a test instance with norm zero, whereas the maximum range corresponds to a test instance with norm equal to $\frac{1}{2}$ (the maximum possible value).  

\subsubsection{Measurements and distance estimate(s)}
\label{subsubsec:meas-and-dist-estimates}
After the \textit{Bell-H} operation, the state of the qubits shown in \cref{eq:final-delta-state}, i.e., the first qubit in the circuit and the index register $\ket{j}$, is measured. In particular, the first qubit is measured first. As a consequence, when 1 (0) is obtained, the indices of the nearest neighbors will have the highest (lowest) probability. If the index register were measured first, the indices would be uniformly sampled. By iterating the circuit execution and the measurement process, the joint probabilities $P(0,j)$ and $P(1,j)$ are estimated as relative frequencies, allowing in turn the estimation of the Euclidean distances. Indeed, the following relationships hold:
\begin{align}
    P(0, j) = \frac{1 + \langle \mathvec{x}_j, \mathvec{x}'_j\rangle}{2N} \implies \langle \mathvec{x}_j, \mathvec{x}'_j\rangle = 2N \times P(0, j) - 1, \label{eq:joint-p0} \\
    P(1, j) = \frac{1 - \langle \mathvec{x}_j, \mathvec{x}'_j\rangle}{2N} \implies \langle \mathvec{x}_j, \mathvec{x}'_j\rangle = 1 - 2N \times P(1, j). \label{eq:joint-p1}
\end{align}
Moreover, for the \textit{extension} encoding (see \cref{tab:encodings-properties}),
\begin{equation}
    d(\mathvec{v}_j, \mathvec{v}') = \sqrt{\frac{3}{4} \langle \mathvec{x}_j, \mathvec{x}'_j\rangle + \|\mathvec{v}'\|^2},
    \label{eq:extension-eucl-dist}
\end{equation}
where $d(\mathvec{v}_j, \mathvec{v}')$ is the Euclidean distance between $\mathvec{v}_j$ and $\mathvec{v}'$, while, for the \textit{translation} encoding,
\begin{equation}
    d(\mathvec{v}_j, \mathvec{v}') = \sqrt{\langle \mathvec{x}_j, \mathvec{x}'_j\rangle + \frac{1}{4} + \|\mathvec{v}'\|^2}.
    \label{eq:translation-eucl-dist}
\end{equation}

Regardless of the encoding used, the Euclidean distances $d(\mathvec{v}_j, \mathvec{v}')$ can be estimated from either $P(0, j)$ or $P(1, j)$. In this work, two ways of combining the information of the two joint probabilities have been devised and tested: \textit{avg} and \textit{diff}. In detail, the \textit{avg} distance estimate is the average of the Euclidean distance estimated from $P(0, j)$ and the Euclidean distance estimated from $P(1, j)$. Instead, for the \textit{diff} distance estimate, the scalar product value is obtained as
\begin{equation*}
    \langle \mathvec{x}_j, \mathvec{x}'_j\rangle = N \times (P(0,j) - P(1,j)),
    \label{eq:diff-scalar-prod}
\end{equation*}
and the Euclidean distance is retrieved through \cref{eq:extension-eucl-dist} or \eqref{eq:translation-eucl-dist}, depending on the encoding used. 

Eventually, it is worth making two observations: given the estimates of $P(0, j)$ and $P(1, j)$ for two indices $j_1, j_2 \in \{0, ..., N-1\}$, the corresponding training instances might be sorted differently according to the \textit{avg} and \textit{diff} distance estimates; if the distance estimates can be mathematically retrieved (i.e., the arguments of the square roots are non-negative), the \textit{avg} distance estimate is always lower than or equal to the corresponding \textit{diff} estimate. More details about these observations can be found in \cref{sec:app-dist-estimates}.

\subsubsection{K nearest neighbors and classification}
\label{subsubsec:k-nn-and-classification}
Once all the Euclidean distances $d(\mathvec{v}_j, \mathvec{v}')$ have been estimated, the training elements are classically sorted according to them. Then, the $k$ nearest neighbors are identified, and the test instance is classified by means of a majority voting on the labels of the nearest neighbors.  

\subsection{Complexity observations}
\label{subsec:complexity-observations}
In terms of complexity, the difference between the quantum $k$-NN algorithm based on the Euclidean distance metric and its classical counterpart lies in the estimation/computation of the Euclidean distances. Indeed, the data preprocessing, the $k$ nearest neighbors identification (performed through a distance sorting operation), and the classification are done classically in both cases. 

For the classical $k$-NN algorithm, the complexity of the Euclidean distances computation is $O(Nd)$. Instead, in the quantum $k$-NN algorithm, the Euclidean distances are not computed exactly but are estimated. In order to do this, a certain number of \textit{shots}, namely, measurements, iterations, is needed. More in detail, each iteration requires to prepare the initial state, run the \textit{Bell-H} quantum circuit, and measure the state of the qubits. The initial state preparation is a quite complex operation. If a QRAM is available, assuming that the real values $x_{ji}$ and $x'_{ji}$ are stored in the QRAM as classical floating point numbers, the initial state can be prepared with complexity $O(\log (NF))$. Indeed, the states $\ket{\alpha}$ and $\ket{\beta}$ can be retrieved from the QRAM with complexity $O(\log (NF))$. Instead, if a QRAM is not available, it is necessary to prepare the desired state starting from $\ket{0}^{\otimes (1+I)}$, and the number of gates needed depends on the architecture of the quantum processor. Regarding the \textit{Bell-H} quantum circuit, it consists of a constant number of elementary gates, therefore its complexity is $O(1)$. The measurement step also has constant complexity, since the measurements are performed in parallel. Eventually, given the state counts, the computation of the distance estimates has cost $O(N)$. Hence, if a QRAM is available, the overall complexity is $O(\mathit{shots} \times \log (NF) + N)$, which is equal to $O(\mathit{shots} \times (\log N + \log d) + N)$. It is also worth highlighting that the higher $N$, the higher the number of shots needed to estimate the Euclidean distances. In conclusion, if we assume that \textit{shots} is a fixed value, the complexity of the Euclidean distances estimation in the quantum $k$-NN algorithm turns out to be $O(\log d + N)$, which is lower than $O(Nd)$. Instead, if we assume that \textit{shots} depends logarithmically on $N$, the complexity turns out to be $O(\log N \log d + N)$, which is still lower than $O(Nd)$.

\section{Implementation}
\label{sec:implementation}
\begin{figure}[htb]
    \centering
    \includegraphics[width=0.85\linewidth]{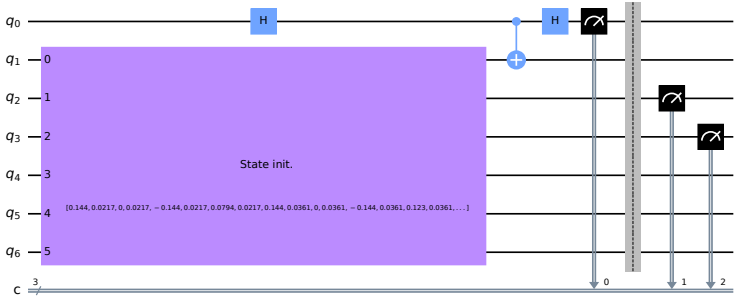}
    \caption{Example of quantum circuit for the quantum $k$-NN based on the Euclidean distance. In detail, $N=4$, $d=2$, and the execution modality is \textit{simulation} (\textit{statevector} does not include the final measurements)}
    \label{fig:eucl-qknn-sample-circuit}
\end{figure}

This section deals with the implementation of the algorithm presented in \cref{sec:method}. In detail, the Euclidean distance quantum $k$-NN has been implemented in Python using Qiskit, the open-source SDK provided by IBM \citep{qiskit}. The code, which is publicly available at \url{https://github.com/ZarHenry96/euclidean-quantum-k-nn}, supports different execution modalities, among which: 
\begin{itemize}
    \item \textit{classical}, which does not involve quantum circuits but runs a classical $k$-NN with the Euclidean distance metric, after the preprocessing step described in \cref{subsubsec:data-preprocessing};
    \item \textit{statevector}, which processes the final state vector of the circuit and, in actual fact, represents an ideal execution with infinite iterations (in this case, no measurement is performed);
    \item \textit{simulation} (\textit{local simulation} in the code), which samples from the final probability distribution of the circuit in order to provide state counts.
\end{itemize}
None of these modalities takes into account the presence of noise. Furthermore, a sample circuit (for the \textit{simulation} modality) is shown in \cref{fig:eucl-qknn-sample-circuit}.

In general, the implementation of the algorithm adheres to the description provided in \cref{subsec:algorithm}, but some technical aspects deserve to be mentioned. Concerning the preprocessing step, each feature is normalized by: subtracting the average of the maximum and the minimum feature values in the training set; dividing by the feature range (computed on the training set, and set to 1 if the feature is constant) multiplied by $\sqrt{d}$. In addition, if a feature of the test instance exceeds the target range after the normalization, it is clipped to the exceeded edge value. Then, let us focus on the execution modalities involving quantum circuits, since the functioning of the \textit{classical} one is quite straightforward. Specifically, the preparation of the initial state $\ket{\psi}$ (Equation \ref{eq:initial-psi-state}) is done by providing the initialization function supplied by Qiskit with the amplitudes of all qubits except the first one, which is in state $\ket{0}$ by default. As regards the indices not associated with any training instance, which exist if $N$ is not a power of two, they are excluded when computing the distance values (hence, they are kept in the joint probabilities estimation). In addition, if the argument of the square root in \cref{eq:extension-eucl-dist} or \eqref{eq:translation-eucl-dist} is negative or larger than one due to the state counts distribution obtained, the distance value is approximated to zero and one, respectively. Eventually, the training instances with the same Euclidean distance are sorted by increasing index in the training set (this holds also for the \textit{classical} execution modality).    

To conclude, it is worth highlighting that the Laplace smoothing \citep{laplace_smoothing} has been applied to the estimation of the joint probabilities in the \textit{simulation} modality. In practice, given a number of counts $c$ for the state $\ket{a}\ket{j}$, with $a \in \{0, 1\}$, the probability $P(a, j)$ is estimated as
\begin{equation*}
    P(a, j) = \frac{c + p}{\mathit{shots} + 2Np}, 
    \label{eq:laplace-smoothing-p0j}
\end{equation*}
where $p$ is the number of pseudocounts added (for each state), and $\mathit{shots}$ is the total number of measurements. In detail, the pseudocounts are summed only to the counts of the significant indices, i.e., the indices actually associated with training instances.

\section{Empirical evaluation}
\label{sec:empirical-evaluation}
In this section, the methods tested, the datasets used, the experimental setup employed, and the results obtained are presented. In particular, the experiments have been run on a shared machine with an Intel Xeon Gold 6238R processor running at 2.20GHz and 125 GB of RAM.

\subsection{Methods}
\label{subsec:methods}
The quantum $k$-NN based on the Euclidean distance introduced in \cref{sec:method} has been tested with different execution modalities and under different (\textit{encoding}, \textit{distance estimate}) configurations, which are reported in \cref{tab:methods}. Runs on real quantum devices have not been performed due to the lack of free-access devices with enough qubits. In addition, for comparison, some classical baseline methods (listed in the same table) have been considered; in particular, the results data for these ones have been fetched from the article by \citet{Zardini2022} \citep[more precisely, from][]{Zardini2023_QML_pipeline_results}.

\begin{table}[htb]
    \centering
    \caption{Methods tested}
    \label{tab:methods}
    \vspace{3pt}
    \begin{tabular}{c|c|c}
        \multicolumn{3}{c}{\textbf{Quantum $k$-NN with Euclidean distance}}                    \\ \hline 
        \textbf{Execution modality}  &  \textbf{Encoding}       &  \textbf{Distance estimate}  \\ \hline
        classical                    &  -                       &  -                           \\ \hline
        statevector                  &  extension, translation  &  avg, diff                   \\ \hline
        simulation                   &  extension, translation  &  avg, diff                   \\
    \end{tabular}
    \\ \vspace{8pt}
    \begin{tabular}{c}
        \textbf{Baseline methods} \\ \hline
        $k$-NN with cosine distance \\ \hline
        random forest (100 trees) \\ \hline
        SVM with \{Gaussian, linear\} kernel \\
    \end{tabular}
\end{table}

\subsection{Datasets}
\label{subsec:datasets}
The datasets used in all the experiments have been taken from the article by \citet{Zardini2022} \citep[more precisely, from][]{Zardini2023_QML_pipeline_datasets}, mainly for the comparability of the results with the baseline methods. Specifically, the properties of these datasets are reported in \cref{tab:datasets}. As explained in the aforementioned article, the original versions of the datasets have been picked from the UCI Machine Learning Repository \citep{uci_ml_repo} according to specific criteria, such as numerical features. Then, they have been preprocessed in order to meet precise requirements, like binary class labels. The datasets used here, which are available together with the code at \url{https://github.com/ZarHenry96/euclidean-quantum-k-nn}, are the ones obtained after the reduction of the number of classes, without subsampling. Additional information about the selection criteria and the preprocessing procedure can be found in the original article.    

\begin{table}[htb]
    \hypersetup{hidelinks}
    \centering
    \caption{Properties of the datasets used. Note that the dataset names are links leading to the UCI pages of the original versions of the datasets}
    \label{tab:datasets}
    \begin{tabular}{c|c|c|c}
        \textbf{Name}                                                                                                          & \textbf{Classes} & \textbf{Size} & \textbf{Features} \\ \hline
        \href{https://archive.ics.uci.edu/ml/datasets/Iris}{01\_iris\_setosa\_versicolor}                                      & 2                & 100           & 4                 \\ \hline
        \href{https://archive.ics.uci.edu/ml/datasets/Iris}{01\_iris\_setosa\_virginica}                                       & 2                & 100           & 4                 \\ \hline
        \href{https://archive.ics.uci.edu/ml/datasets/Iris}{01\_iris\_versicolor\_virginica}                                   & 2                & 100           & 4                 \\ \hline
        \href{https://archive.ics.uci.edu/ml/datasets/Blood+Transfusion+Service+Center}{02\_transfusion} \citep{transfusion}   & 2                & 748           & 4                 \\ \hline
        \href{https://archive.ics.uci.edu/ml/datasets/Vertebral+Column}{03\_vertebral\_column\_2C}                             & 2                & 310           & 6                 \\ \hline
        \href{https://archive.ics.uci.edu/ml/datasets/seeds}{04\_seeds\_1\_2}                                                  & 2                & 140           & 7                 \\ \hline
        \href{https://archive.ics.uci.edu/ml/datasets/Ecoli}{05\_ecoli\_cp\_im}                                                & 2                & 220           & 7                 \\ \hline
        \href{https://archive.ics.uci.edu/ml/datasets/Glass+Identification}{06\_glasses\_1\_2}                                 & 2                & 146           & 9                 \\ \hline
        \href{https://archive.ics.uci.edu/ml/datasets/Breast+Tissue}{07\_breast\_tissue\_adi\_fadmasgla}                       & 2                & 71            & 9                 \\ \hline
        \href{https://archive.ics.uci.edu/ml/datasets/Breast+Cancer+Coimbra}{08\_breast\_cancer} \citep{breast_cancer_coimbra} & 2                & 116           & 9                 \\ \hline
        \href{https://archive.ics.uci.edu/ml/datasets/Speaker+Accent+Recognition}{09\_accent\_recognition\_uk\_us}             & 2                & 210           & 12                \\ \hline
        \href{https://archive.ics.uci.edu/ml/datasets/Leaf}{10\_leaf\_11\_9} \citep{leaf}                                      & 2                & 30            & 14                \\
    \end{tabular}
\end{table}

\subsection{Experimental setup}
\label{subsec:experimental-setup}
In all experiments, the stratified $k$-fold cross-validation has been adopted as the validation technique. In practice, each dataset is split into $k$ folds, i.e., subsets. Then, $k-1$ folds form the training set, whereas the leftover represents the test set, and this last step is executed $k$ times so that each subset is used once as the test set. The adjective ``stratified" implies that the ratio between classes in the folds has been kept as close as possible to the one of the given dataset. In addition, the same seed has been used for the folds generation in all experiments; in this way, all methods have been evaluated on the same folds.

The parameters values employed in the quantum $k$-NN experiments are reported in \cref{tab:exp-params}. In particular, for all execution modalities, the number of folds in the $k$-fold cross-validation (\textit{folds}) has been set to 5 (a common value in \ac{ML}), and four different numbers of nearest neighbors selected (\textit{k}) have been considered. Concerning the \textit{simulation} modality, all (\textit{encoding}, \textit{distance estimate}) configurations have been tested with 1024 measurements (\textit{shots}), which is the default value provided by Qiskit, and the best one has been evaluated also varying this parameter value. In addition, the number of pseudocounts for the Laplace smoothing has been arbitrarily set to 10, and 5 runs with different seeds have been performed in order to gain statistical evidence. Specifically, the simulation seed for each test instance is randomly generated starting from a ``root" run seed. Eventually, it is worth highlighting that the different $k$ values have been evaluated on different seeds, while the \textit{avg} and \textit{diff} distance estimates have been evaluated on the same seeds (namely, for each test instance, the two distance estimates are computed using the same state counts). 

\begin{table}[htb]
    \begingroup
        \renewcommand{\thefootnote}{\alph{footnote}}
        \begin{center}
            \caption{Parameters setting for the quantum $k$-NN experiments}
            \label{tab:exp-params}
            \begin{tabular}{c|c}
                \multicolumn{2}{c}{\textbf{Common parameters}}  \\ \hline
                folds                 & 5                       \\ \hline
                k                     & 3, 5, 7, 9              \\
                \multicolumn{2}{c}{}                            \\
            \end{tabular}
            \qquad
            \begin{tabular}{c|c}
                \multicolumn{2}{c}{\textbf{Simulation parameters}}                                                                  \\ \hline
                shots                 & 512\footnotemark[1], 1024, 2048\footnotemark[1], 4096\footnotemark[1], 8192\footnotemark[1] \\ \hline
                pseudocounts          & 10                                                                                          \\ \hline
                runs                  & 5                                                                                           \\
            \end{tabular}
        \end{center}
        \footnotetext{\textsuperscript{a}Only the best (\textit{encoding}, \textit{distance estimate}) configuration of the quantum $k$-NN has been tested with this number of shots.}
    \endgroup
\end{table} 

Regarding the baseline methods considered for comparison \citep{Zardini2022}, the normalization procedure applied to the input data features is a canonical min-max normalization, whose output range is $[0, 1]$. The number of folds, the folds generation seed, and the $k$ values considered for the classical $k$-NN with cosine distance are the same as those used in the quantum $k$-NN experiments. Eventually, the number of runs for the random forest is 5 (it is a stochastic method).

\subsection{Results}
\label{subsec:results}
The results are presented by means of scatterplots and boxplots. In particular, the quantum $k$-NN has been evaluated in terms of classification accuracy and correctness of the nearest neighbors found, whereas, for the baseline methods, only the classification accuracy has been considered. More in detail, given a fold, the accuracy is defined as
\begin{equation*}
    \mathit{accuracy} = \frac{\mathit{number\ of\ correctly\ classified\ instances \ in\ the\ fold}}{\mathit{total\ number\ of\ instances\ in\ the\ fold}}\,;
    \label{eq:accuracy}
\end{equation*}
in the case of multiple runs, the average value across runs is reported. Instead, regarding the correctness of the nearest neighbors found, the Jaccard index and the Average Jaccard score \citep{averageJaccardScore} have been taken into account. Specifically, given a test instance, the Jaccard index (JI) is defined as 
\begin{equation*}
    \mathit{Jaccard\ index\ (JI)} = \frac{\lvert \mathset{S}_c \cap \mathset{S}_f \rvert}{\lvert \mathset{S}_c \cup \mathset{S}_f \rvert}\,,
    \label{eq:jaccard-index}
\end{equation*}
where $\mathset{S}_c$ is the set of correct nearest neighbors (classically computed), and $\mathset{S}_f$ is the set of nearest neighbors found. Since there is a Jaccard index value for each test instance, the average value has been considered for each fold, and the average of this average value across runs is reported. The same is done for the Average Jaccard (AJ) score, which, given a test instance, is defined as 
\begin{equation*}
    \mathit{Average\ Jaccard\ (AJ)} = \frac{1}{k}\sum_{m=1}^{k}\mathit{h}(\mathset{S}_{cm}, \mathset{S}_{fm})\,,
    \label{eq:avg-jaccard-score}
\end{equation*}
where $k$ is the number of nearest neighbors selected, $\mathit{h}$ is the function computing the Jaccard index, and $\mathset{S}_{cm}$ is the set containing the correct nearest neighbors up to the $m$-th most similar item (the same for $\mathset{S}_{fm}$).
Eventually, the statistical significance of the results obtained has been assessed through the Wilcoxon signed-rank test \citep{wilcoxon_test}, as the data are paired; in some cases (difference boxplots), also the one-sample T-test \citep{student_ttest} has been taken into account.

\begin{figure}[b]
    \centering
    \subfloat[\label{fig:cl-vs-sv-acc}]{
        \centering
        \includegraphics[width=0.31\linewidth]{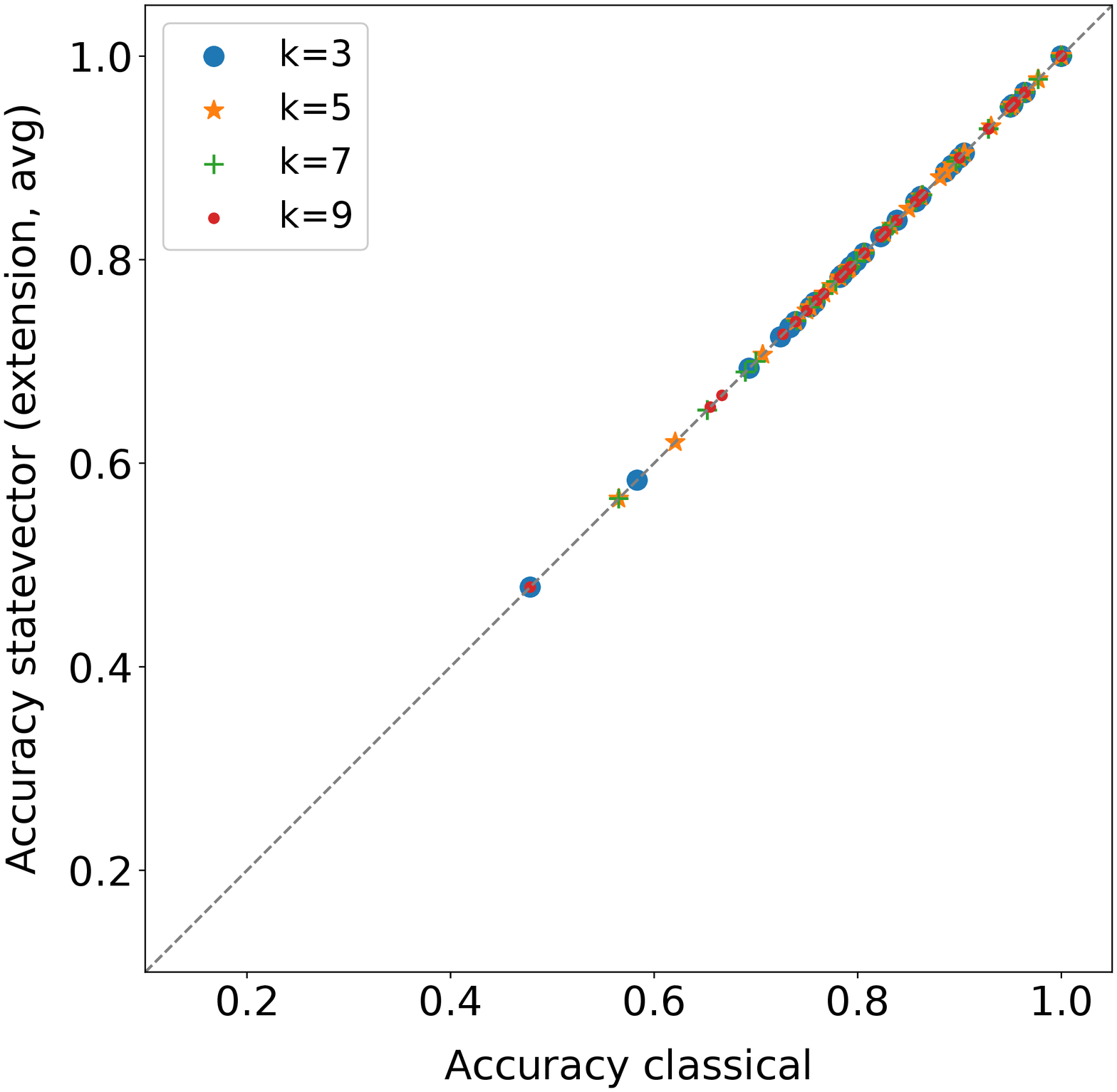}
    }
    \subfloat[\label{fig:cl-vs-sv-ji}]{
        \centering
        \includegraphics[width=0.31\linewidth]{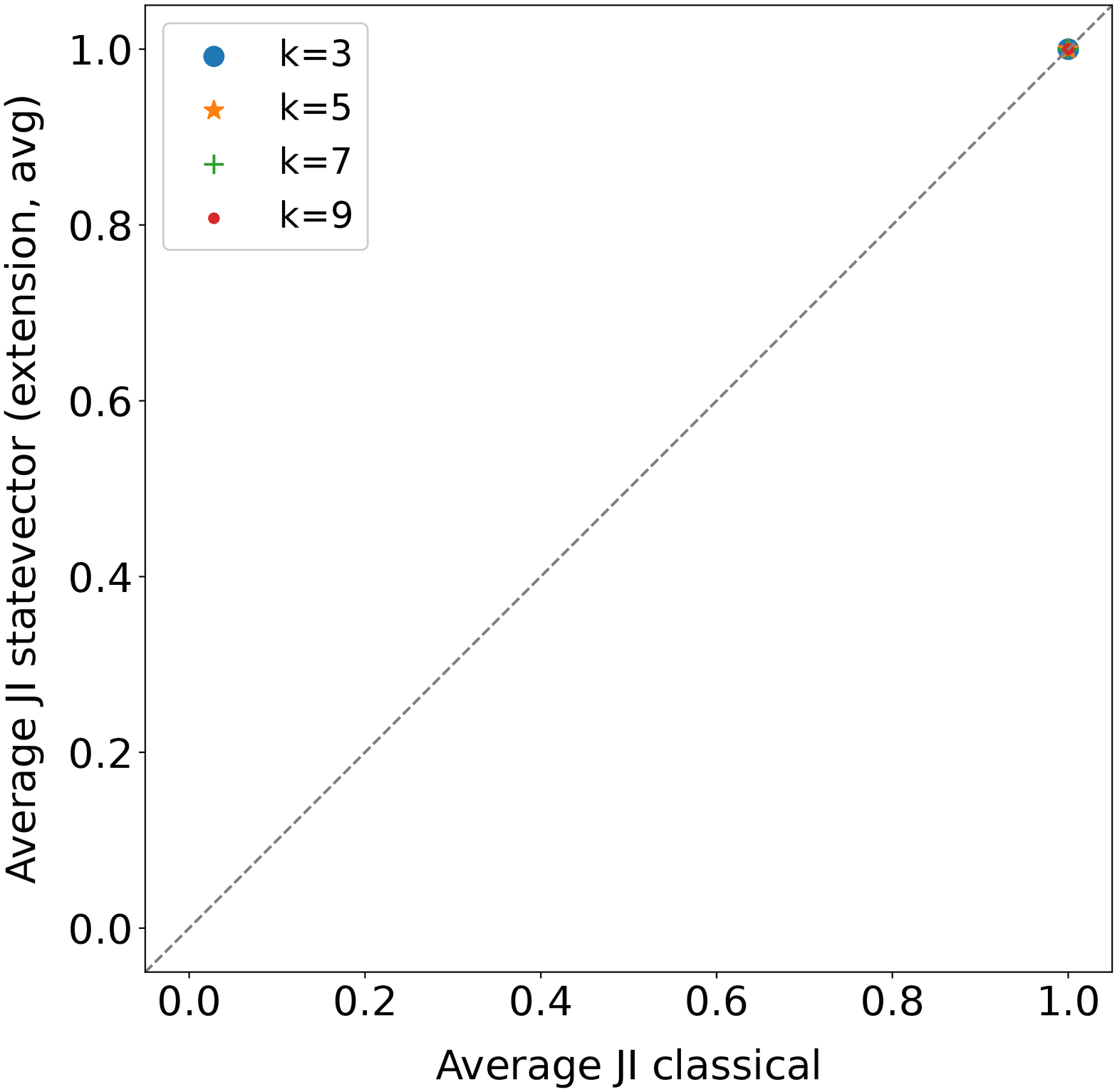}
    }
    \subfloat[\label{fig:cl-vs-sv-aj}]{
        \centering
        \includegraphics[width=0.31\linewidth]{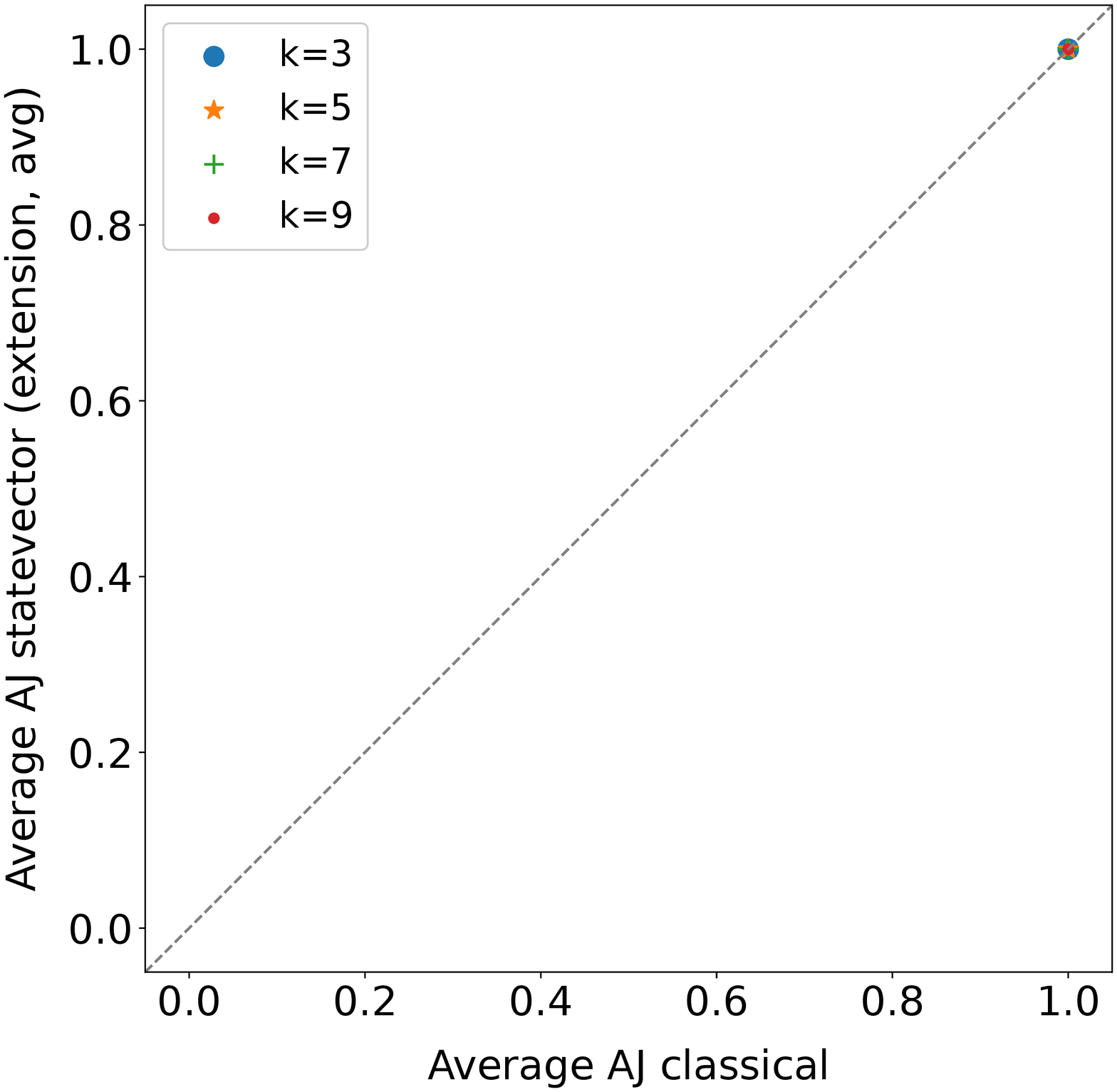}
    }
    \caption{Comparison between \textit{classical} and \textit{statevector} execution modalities in terms of accuracy (a), Jaccard index (b), and Average Jaccard score (c). The configuration used for \textit{statevector} is (\textit{extension}, \textit{avg}), but the results are the same for all configurations. Each point is related to a dataset fold}
    \label{fig:cl-vs-sv}

    \vspace{10pt}

    \hypersetup{hidelinks}
    \centering
    \captionof{table}{Wilcoxon signed-rank test ($\alpha\,{=}\,0.05$) applied to the distributions shown in \cref{fig:cl-vs-sv}. The values reported in the table are the p-values obtained}
    \label{tab:cl-vs-sv-stats}
    \small
    \begin{tabular}{c|c|c|c|c}
                                & \textbf{k=3} & \textbf{k=5} & \textbf{k=7} & \textbf{k=9} \\ \hline
		\cref{fig:cl-vs-sv-acc} & 1.000        & 1.000        & 1.000        & 1.000        \\ \hline
		\cref{fig:cl-vs-sv-ji}  & 1.000        & 1.000        & 1.000        & 1.000        \\ \hline
		\cref{fig:cl-vs-sv-aj}  & 1.000        & 1.000        & 1.000        & 1.000        \\
    \end{tabular}
\end{figure}

\begin{figure}[t!]
    \centering
    \subfloat[\label{fig:sv-vs-si-ext-avg-acc}]{
        \centering
        \includegraphics[width=0.31\linewidth]{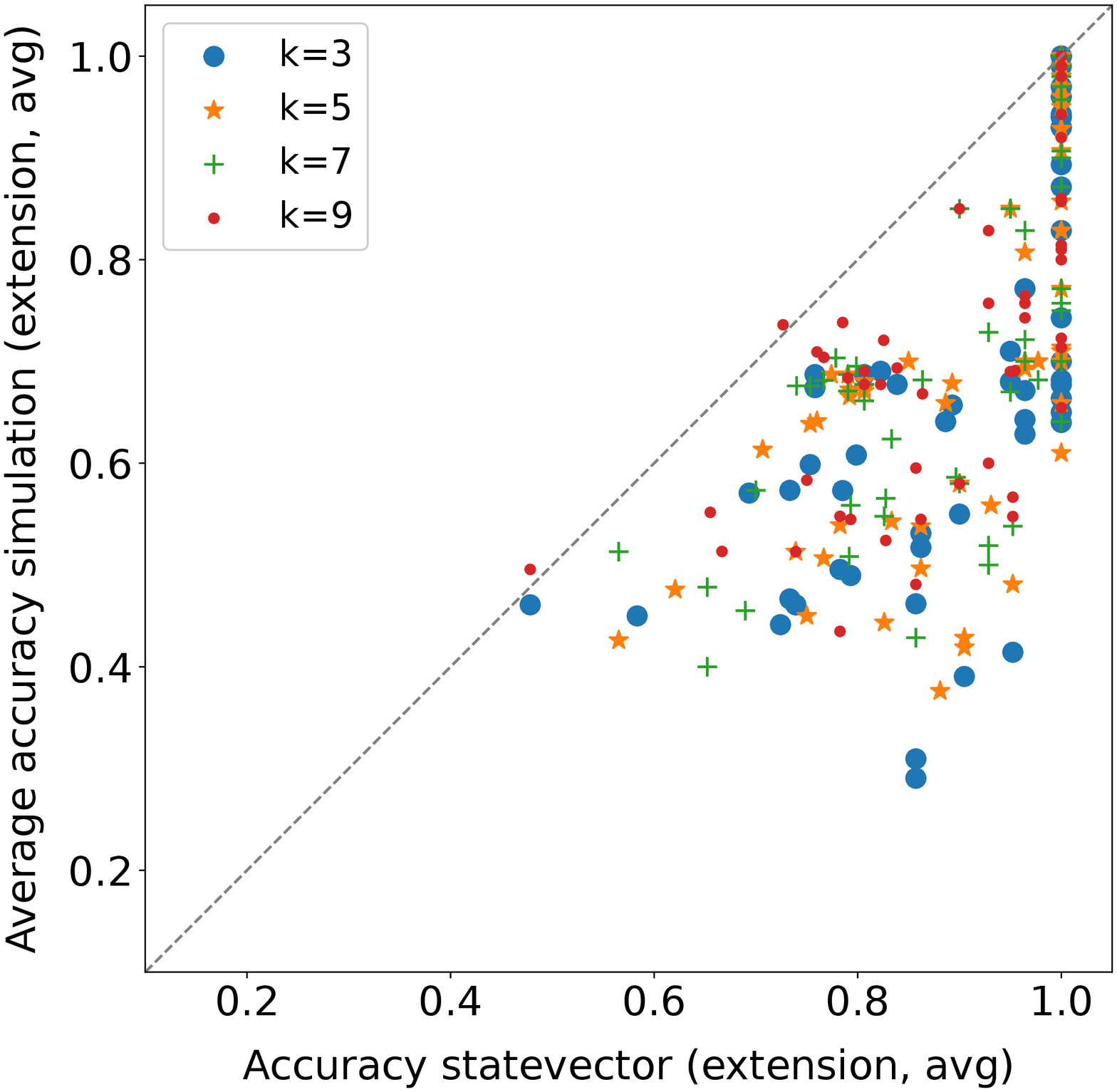}
    }
    \subfloat[\label{fig:sv-vs-si-ext-avg-ji}]{
        \centering
        \includegraphics[width=0.31\linewidth]{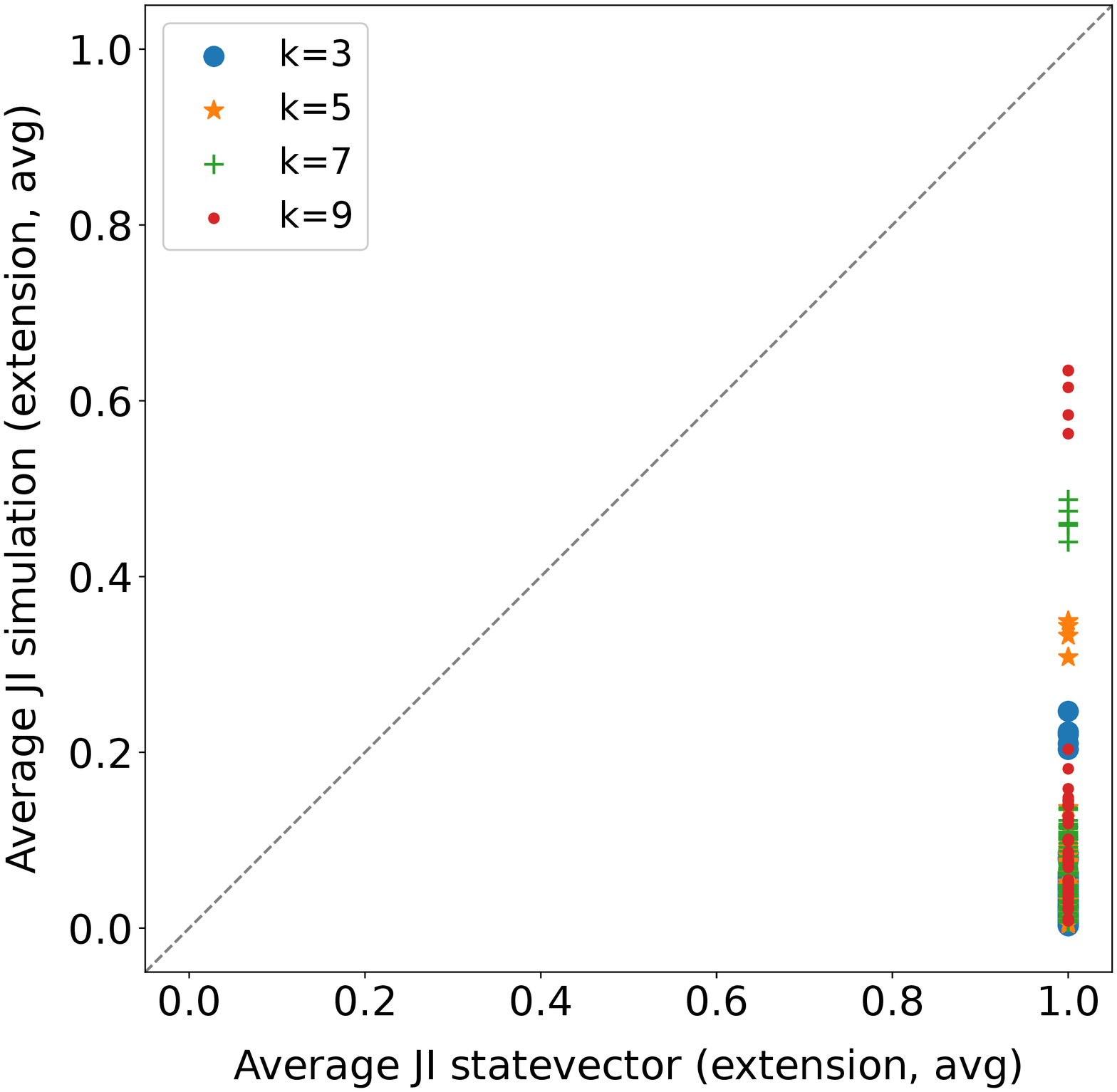}
    }
    \subfloat[\label{fig:sv-vs-si-ext-avg-aj}]{
        \centering
        \includegraphics[width=0.31\linewidth]{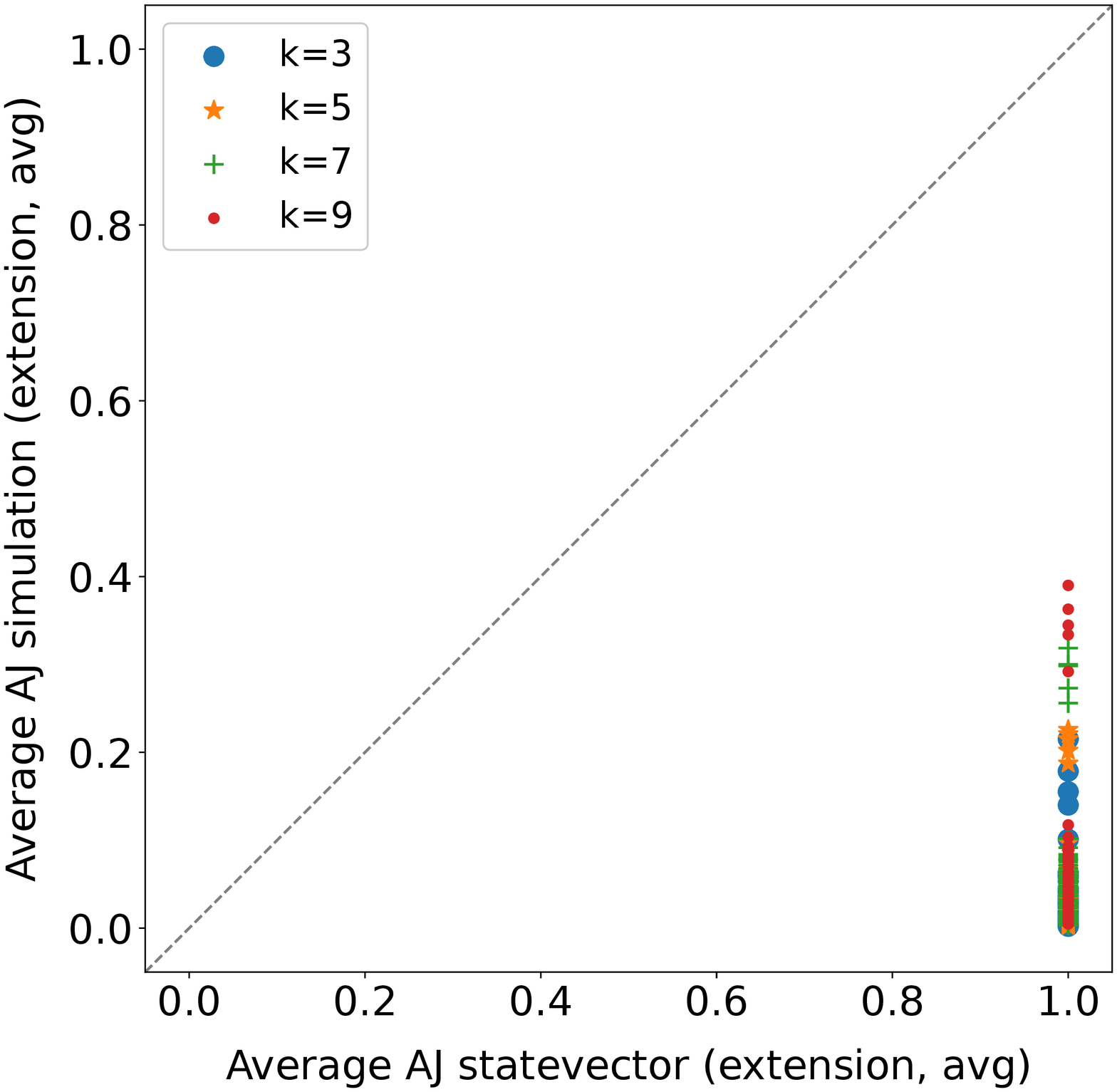}
    }
    \caption{Comparison between \textit{statevector} (\textit{extension}, \textit{avg}) and \textit{simulation} (\textit{extension}, \textit{avg}) in terms of accuracy (a), Jaccard index (b), and Average Jaccard score (c). The number of shots for \textit{simulation} is 1024, and each point is related to a dataset fold}
    \label{fig:sv-vs-si-ext-avg}

    \vspace{10pt}

    \hypersetup{hidelinks}
    \centering
    \captionof{table}{Wilcoxon signed-rank test ($\alpha\,{=}\,0.05$) applied to the distributions shown in \cref{fig:sv-vs-si-ext-avg}. The values reported in the table are the p-values obtained}
    \label{tab:sv-vs-si-ext-avg-stats}
    \small
    \begin{tabular}{c|c|c|c|c}
                                        & \textbf{k=3} & \textbf{k=5} & \textbf{k=7} & \textbf{k=9} \\ \hline
		\cref{fig:sv-vs-si-ext-avg-acc} & 1.624E-10    & 1.622E-10    & 5.140E-10    & 1.608E-09    \\ \hline
		\cref{fig:sv-vs-si-ext-avg-ji}  & 1.626E-11    & 1.630E-11    & 1.629E-11    & 1.630E-11    \\ \hline
		\cref{fig:sv-vs-si-ext-avg-aj}  & 1.629E-11    & 1.630E-11    & 1.630E-11    & 1.630E-11    \\
    \end{tabular}
\end{figure}

\subsubsection{Execution modalities comparison}
\label{subsubsec:exec-modalities-comp}
Let us consider first the \textit{classical} and \textit{statevector} execution modalities. As shown in \cref{fig:cl-vs-sv}, the two modalities are equivalent in terms of accuracy (\cref{fig:cl-vs-sv-acc}), Jaccard index (\cref{fig:cl-vs-sv-ji}), and Average Jaccard score (\cref{fig:cl-vs-sv-aj}); the absence of a statistical difference is certified by the Wilcoxon signed-rank test (\cref{tab:cl-vs-sv-stats}). Only one \textit{statevector} configuration, i.e., (\textit{extension}, \textit{avg}), is shown here, but the results are identical for all of them\footnote{For the \textit{translation} encoding, in one fold of one dataset, there are cases in which two nearest neighbors are swapped due to the numerical approximation of the distance values. Therefore, the Average Jaccard score for that fold turns out to be slightly lower than that of the \textit{classical}, but the difference is not statistically significant (p-value=$0.317$ for all $k$ values).}. This confirms that the algorithm presented in \cref{sec:method} is correct; indeed, in the ideal case, it is able to achieve the same results as its classical counterpart. It is also worth remembering that the advantage of the quantum algorithm with respect to its classical counterpart lies in the execution time. 

Then, let us focus on the \textit{statevector} and \textit{simulation} execution modalities. Specifically, \cref{fig:sv-vs-si-ext-avg} shows the comparison in terms of accuracy (\cref{fig:sv-vs-si-ext-avg-acc}), Jaccard index (\cref{fig:sv-vs-si-ext-avg-ji}), and Average Jaccard score (\cref{fig:sv-vs-si-ext-avg-aj}) for the (\textit{extension}, \textit{avg}) configuration. As expected, the limited number of measurements (1024) leads to a substantial performance worsening, and \textit{statevector} turns out to statistically outperform the \textit{simulation} modality both in classification accuracy and correctness of the nearest neighbors found, as reported in \cref{tab:sv-vs-si-ext-avg-stats}. In particular, the drop in performance is more marked for the Jaccard index and the Average Jaccard score. These observations hold for all the (\textit{encoding}, \textit{distance estimate}) configurations; analogous plots and related significance tables for the other configurations are available in \cref{subsec:app-exec-modalities-comp}.

\subsubsection{Encodings and distance estimates comparison}
\label{subsubsec:encodings-and-dist-ests-comp}
\begin{figure}[b!]
    \centering
    \subfloat[\label{fig:encodings-and-dist-ests-comp-acc}]{
        \includegraphics[width=0.98\linewidth]{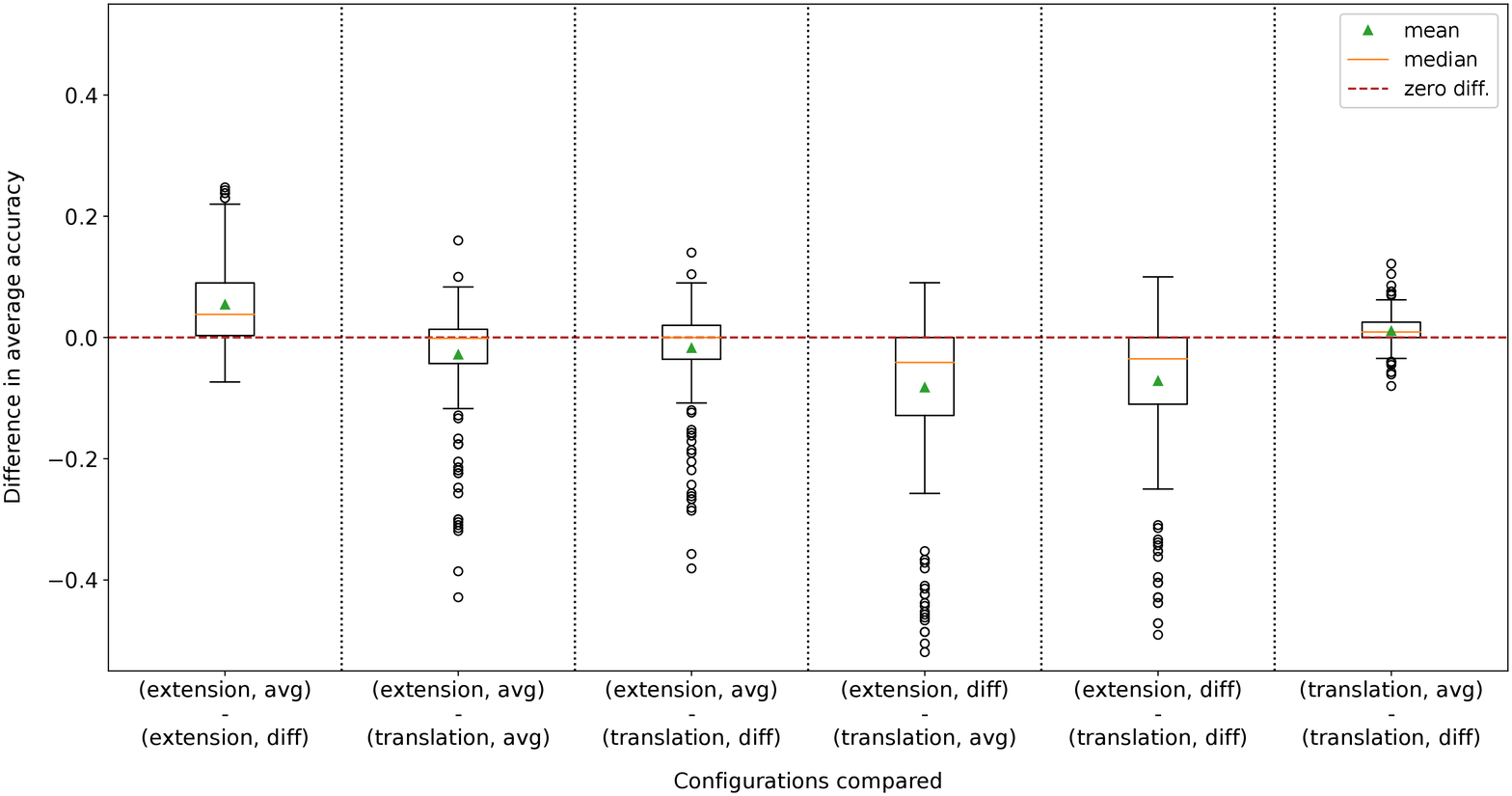}
    } \\
    \subfloat[\label{fig:encodings-and-dist-ests-comp-ji}]{
        \includegraphics[width=0.98\linewidth]{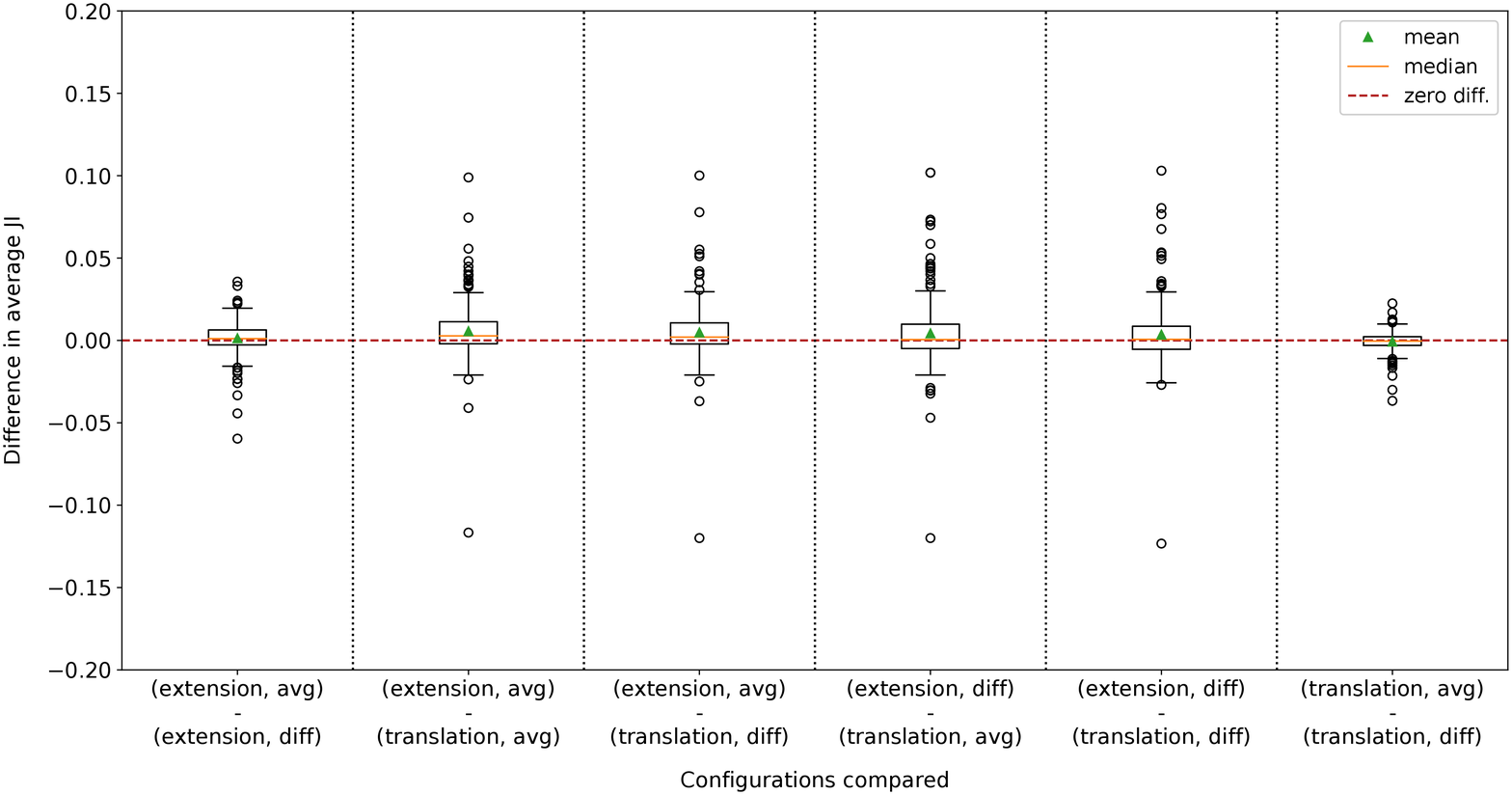}
    }
    \caption{Comparison of (\textit{encoding}, \textit{distance estimate}) configurations in terms of accuracy (a) and Jaccard index (b) for the \textit{simulation} execution modality. The number of shots is 1024, and each data point corresponds to the difference for a (\textit{dataset fold}, \textit{k value}) pair}
    \label{fig:encodings-and-dist-ests-comp}
\end{figure}

Since the various (\textit{encoding}, \textit{distance estimate}) configurations have achieved the same results for the \textit{statevector} execution modality (the Euclidean distance estimates are exact), only \textit{simulation} (with 1024 shots) is taken into account here. In detail, the configurations are compared by means of difference boxplots, in which each data point represents the difference for a (\textit{dataset fold}, \textit{k value}) pair. The comparisons in accuracy and Jaccard index are shown in \cref{fig:encodings-and-dist-ests-comp-acc} and \cref{fig:encodings-and-dist-ests-comp-ji}, respectively, while the plot for the Average Jaccard score is available in \cref{subsec:app-encodings-and-dist-ests-comp} (\cref{fig:encodings-and-dist-ests-comp-aj}) for space reasons.  

Concerning the classification accuracy, the configuration that has achieved the best results is (\textit{translation}, \textit{avg}), which has statistically outperformed all the others, as confirmed by \cref{tab:encodings-and-dist-ests-comp-acc-stats}. In general, the \textit{translation} encoding has performed better than the \textit{extension} encoding in accuracy, and, with an equal encoding, the \textit{avg} distance estimate has outperformed the \textit{diff} distance estimate. Moreover, the differences are statistically significant in terms of both median (Wilcoxon signed-rank test) and mean (one-sample T-test), except for the (\textit{extension}, \textit{avg}) - (\textit{translation}, \textit{diff}) comparison in terms of median.

\begin{table}[t!]
    \hypersetup{hidelinks}
    \centering
    \caption{Wilcoxon signed-rank test and one-sample T-test applied to the distributions shown in \cref{fig:encodings-and-dist-ests-comp-acc} (a) and \cref{fig:encodings-and-dist-ests-comp-ji} (b). Each column corresponds to a different comparison, and the first letter identifies the encoding (E=\textit{extension}, T=\textit{translation}), while the second letter identifies the distance estimate (A=\textit{avg}, D=\textit{diff}). The values reported in the tables are the p-values obtained ($\alpha\,{=}\,0.05$)}
    \label{tab:encodings-and-dist-ests-comp-stats}
    \captionsetup{position=top}
    \subfloat[\label{tab:encodings-and-dist-ests-comp-acc-stats}]{
        \begin{tabular}{c|c|c|c|c|c|c}
                     & \textbf{EA-ED} & \textbf{EA-TA} & \textbf{EA-TD} & \textbf{ED-TA} & \textbf{ED-TD} & \textbf{TA-TD} \\ \hline
    		Wilcoxon & 9.378E-26      & 3.019E-05      & 0.069          & 1.771E-24      & 1.539E-21      & 1.084E-09      \\ \hline
    		T-test   & 1.236E-27      & 3.706E-07      & 0.001          & 1.404E-20      & 2.037E-18      & 2.492E-09      \\
        \end{tabular}
    } \\
    \subfloat[\label{tab:encodings-and-dist-ests-comp-ji-stats}]{
        \begin{tabular}{c|c|c|c|c|c|c}
                     & \textbf{EA-ED} & \textbf{EA-TA} & \textbf{EA-TD} & \textbf{ED-TA} & \textbf{ED-TD} & \textbf{TA-TD} \\ \hline
    		Wilcoxon & 0.003          & 8.888E-09      & 1.917E-07      & 0.010          & 0.043          & 0.104          \\ \hline
    		T-test   & 0.054          & 6.775E-07      & 1.102E-05      & 0.001          & 0.005          & 0.054          \\ 
        \end{tabular}
    }
\end{table}

Surprisingly, the configuration with the highest classification accuracy is not the one that has found the best nearest neighbors. Indeed, the configuration that has achieved the best results in Jaccard index is (\textit{extension}, \textit{avg}), and the differences with respect to the other ones are almost always significant as reported in \cref{tab:encodings-and-dist-ests-comp-ji-stats}. In general, the \textit{extension} encoding has outperformed the \textit{translation} encoding in Jaccard index, while, with an equal encoding, there is not a clear winning distance estimate: the \textit{avg} distance estimate has performed better with the \textit{extension} encoding, whereas the \textit{diff} distance estimate has achieved better results with the \textit{translation} encoding. The differences are almost all significant in terms of both median and mean; the only exceptions are the (\textit{extension}, \textit{avg}) - (\textit{extension}, \textit{diff}) comparison in mean, and the (\textit{translation}, \textit{avg}) - (\textit{translation}, \textit{diff}) comparison for both statistics. Regarding the Average Jaccard score (\cref{fig:encodings-and-dist-ests-comp-aj}), the trend is similar, with (\textit{extension}, \textit{avg}) being the best configuration. However, in this case, the \textit{extension} encoding with the \textit{diff} distance estimate has performed the worst, which means that among the $k$ nearest neighbors selected by this configuration, the correct ones are placed in the last positions. In addition, few differences are statistically significant as reported in \cref{tab:encodings-and-dist-ests-comp-aj-stats}. Among these, it is worth mentioning the (\textit{extension}, \textit{avg}) configuration statistically outperforming all the others in terms of median and (\textit{extension}, \textit{diff}) also in terms of mean.

\subsubsection{Comparison with baseline methods}
\label{subsubsec:comp-with-baselines}
\begin{figure}[t!]
    \centering
    \subfloat[\label{fig:sv-transl-avg-vs-knn-cos}]{
        \centering
        \includegraphics[width=0.33\linewidth]{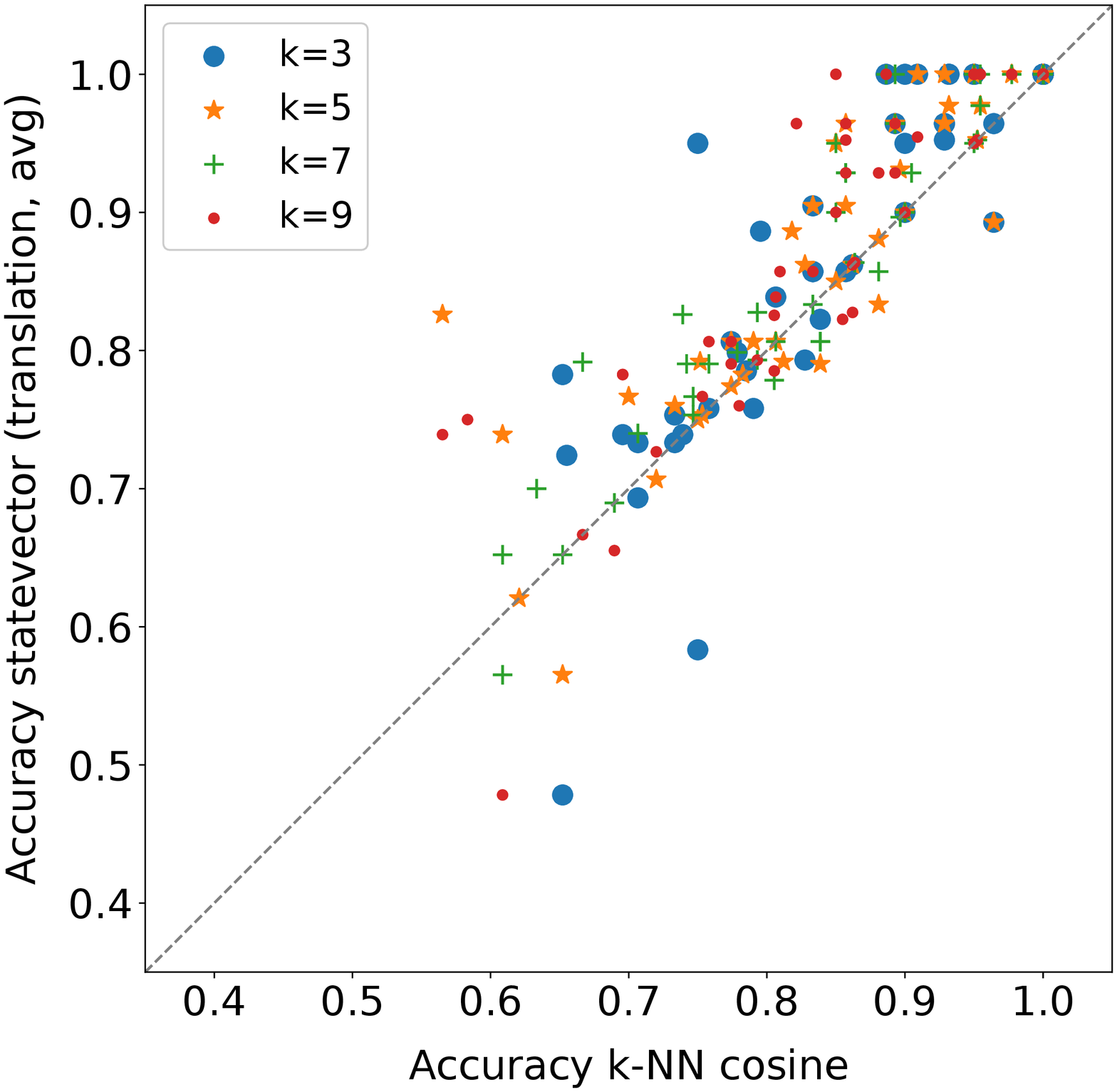}
    }
    \subfloat[\label{fig:sv-transl-avg-vs-random-forest}]{
        \centering
        \includegraphics[width=0.33\linewidth]{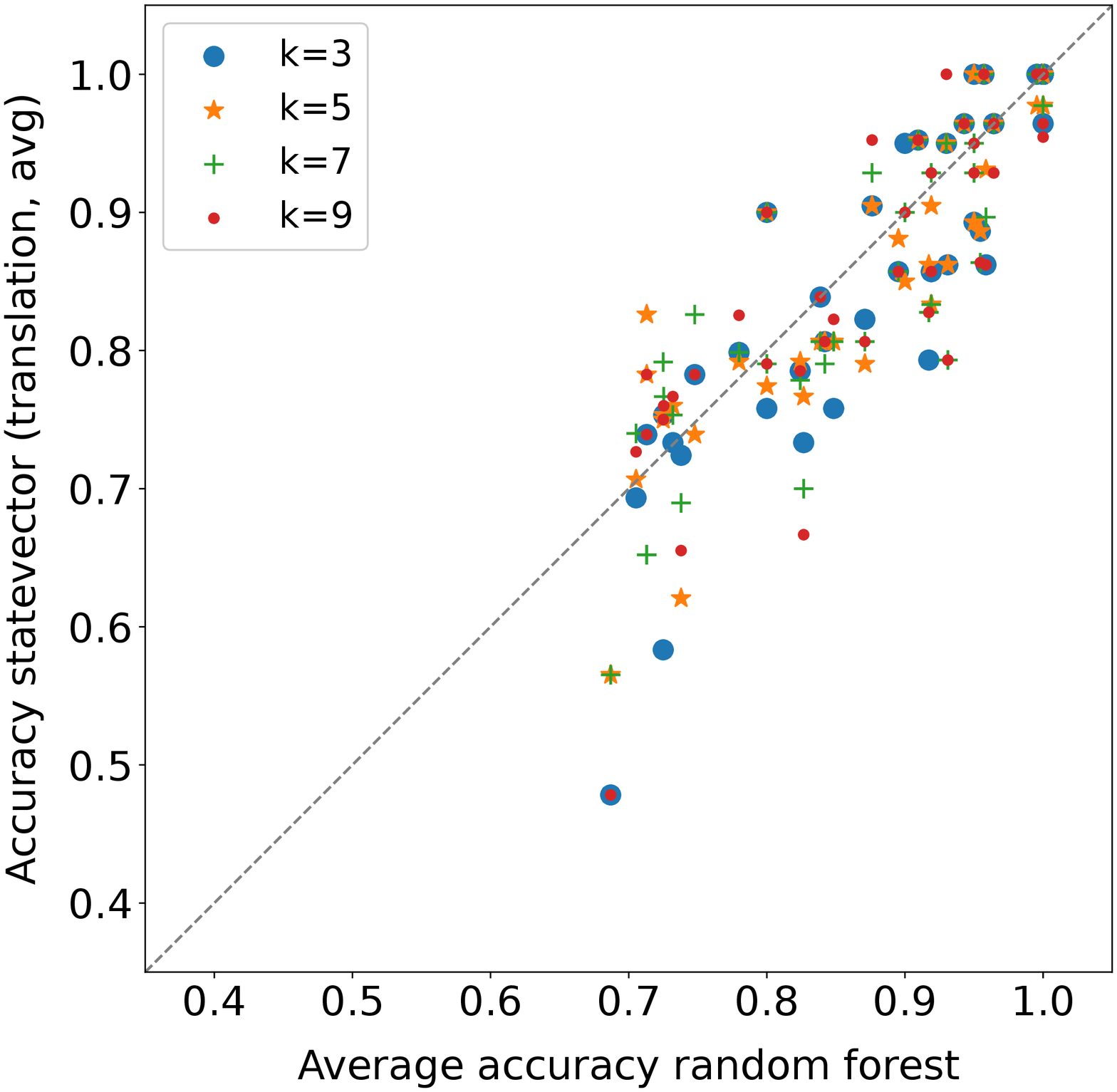}
    } \\
    \subfloat[\label{fig:sv-transl-avg-vs-svm-gaussian}]{
        \centering
        \includegraphics[width=0.33\linewidth]{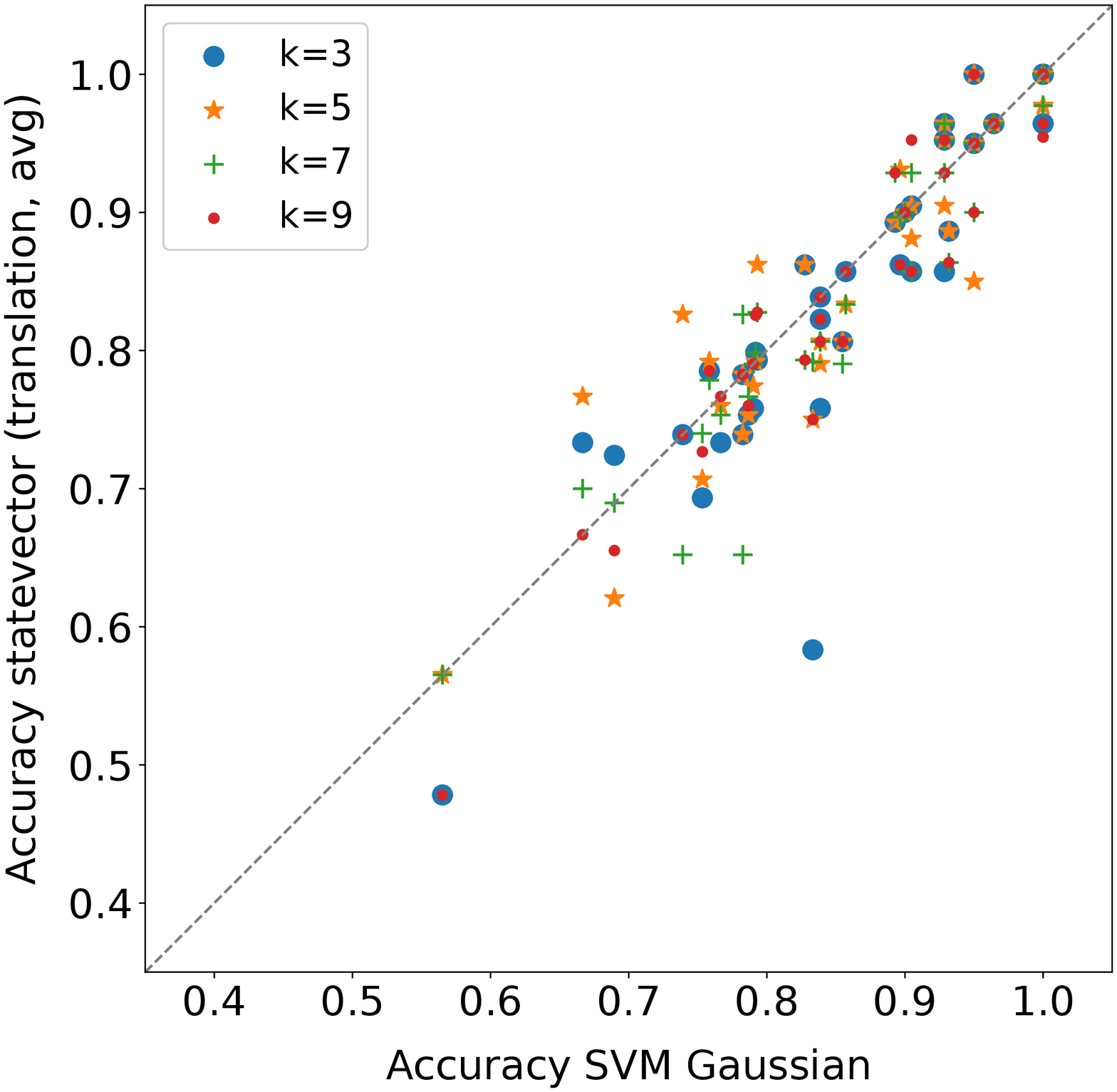}
    }
    \subfloat[\label{fig:sv-transl-avg-vs-svm-linear}]{
        \centering
        \includegraphics[width=0.33\linewidth]{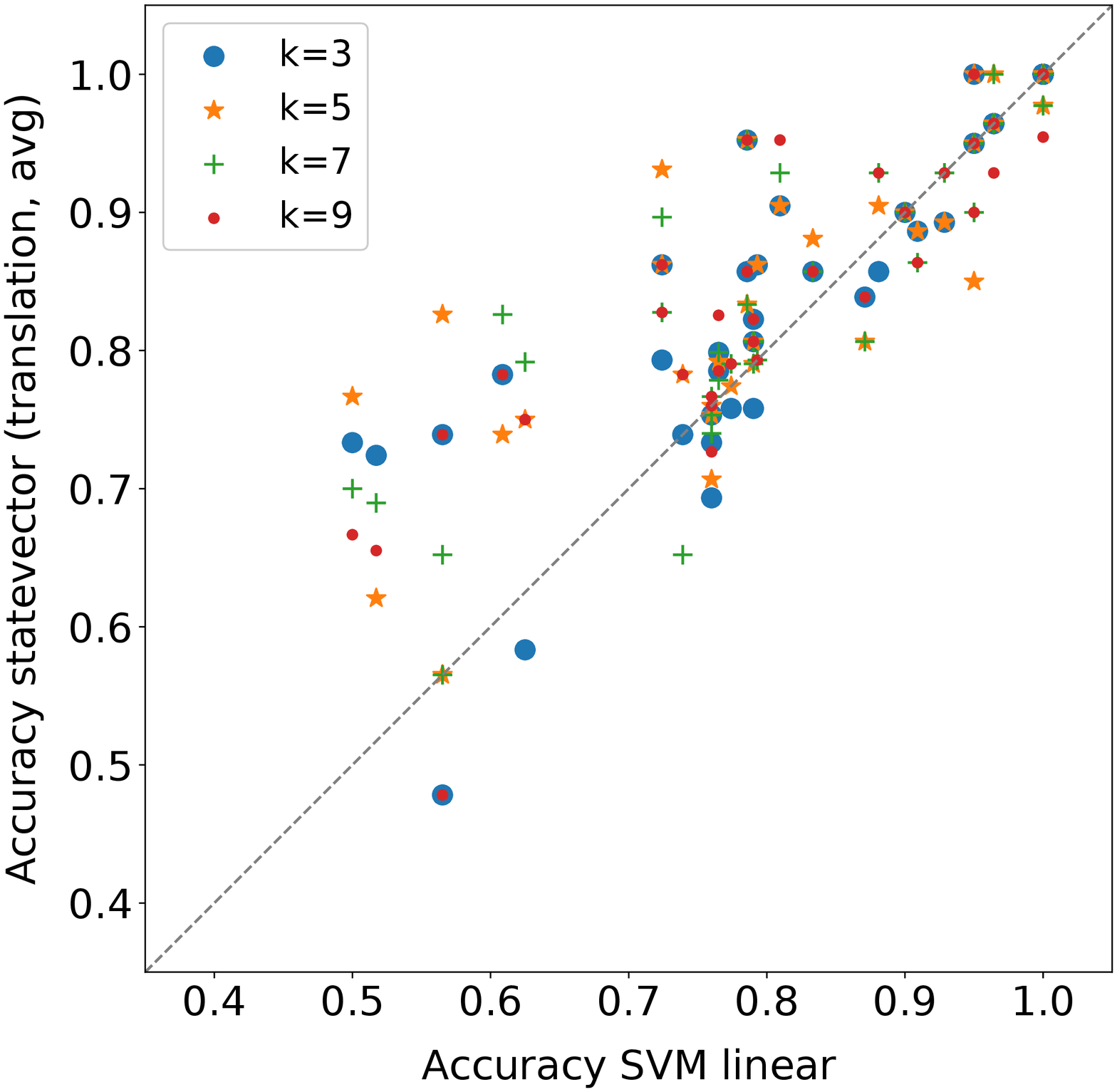}
    }
    \caption{Comparison between some classical baseline methods and \textit{statevector} in terms of accuracy. The configuration used for \textit{statevector} is (\textit{translation}, \textit{avg}), but the results are the same for all configurations. Each point is related to a dataset fold}
    \label{fig:sv-comp-with-baselines}

    \vspace{10pt}

    \hypersetup{hidelinks}
    \centering
    \captionof{table}{Wilcoxon signed-rank test ($\alpha\,{=}\,0.05$) applied to the distributions shown in \cref{fig:sv-comp-with-baselines}. The values reported in the table are the p-values obtained}
    \label{tab:sv-comp-with-baselines-stats}
    \small
    \begin{tabular}{c|c|c|c|c}
                                                  & \textbf{k=3} & \textbf{k=5} & \textbf{k=7} & \textbf{k=9} \\ \hline
		\cref{fig:sv-transl-avg-vs-knn-cos}       & 0.003        & 0.001        & 6.502E-05    & 8.149E-05    \\ \hline
		\cref{fig:sv-transl-avg-vs-random-forest} & 0.074        & 0.165        & 0.095        & 0.258        \\ \hline
		\cref{fig:sv-transl-avg-vs-svm-gaussian}  & 0.046        & 0.407        & 0.103        & 0.062        \\ \hline
        \cref{fig:sv-transl-avg-vs-svm-linear}    & 0.045        & 0.005        & 0.012        & 0.007        \\ 
    \end{tabular}
\end{figure}

Some classical baseline methods have been chosen for comparison in terms of classification accuracy. \cref{fig:sv-comp-with-baselines} shows the comparisons with the \textit{statevector} execution modality in the (\textit{translation}, \textit{avg}) configuration; actually, the configuration used is irrelevant for \textit{statevector}, as explained in the previous sections. In practice, in the ideal case, the quantum $k$-NN based on the Euclidean distance metric statistically outperforms both the classical $k$-NN with the cosine distance metric (\cref{fig:sv-transl-avg-vs-knn-cos}) and the SVM with the linear kernel (\cref{fig:sv-transl-avg-vs-svm-linear}), as confirmed by \cref{tab:sv-comp-with-baselines-stats}. Instead, it is outperformed by both the random forest (\cref{fig:sv-transl-avg-vs-random-forest}) and the SVM with the Gaussian kernel (\cref{fig:sv-transl-avg-vs-svm-gaussian}), although the differences are almost never statistically significant (only the difference with respect to the SVM with the Gaussian kernel, for $k=3$, is significant).

Analogous comparison plots for the \textit{simulation} execution modality in the (\textit{translation}, \textit{avg}) configuration, namely, the configuration that has achieved the best results in classification accuracy, are available in \cref{subsec:app-comp-with-baselines} (\cref{fig:si-comp-with-baselines}). Specifically, all baseline methods considered have statistically outperformed the quantum $k$-NN in the \textit{simulation} execution modality, as confirmed by \cref{tab:si-comp-with-baselines-stats}.

\begin{figure}[t!]
    \centering
    \subfloat[\label{fig:shots-num-comp-acc}]{
        \includegraphics[width=0.48\linewidth]{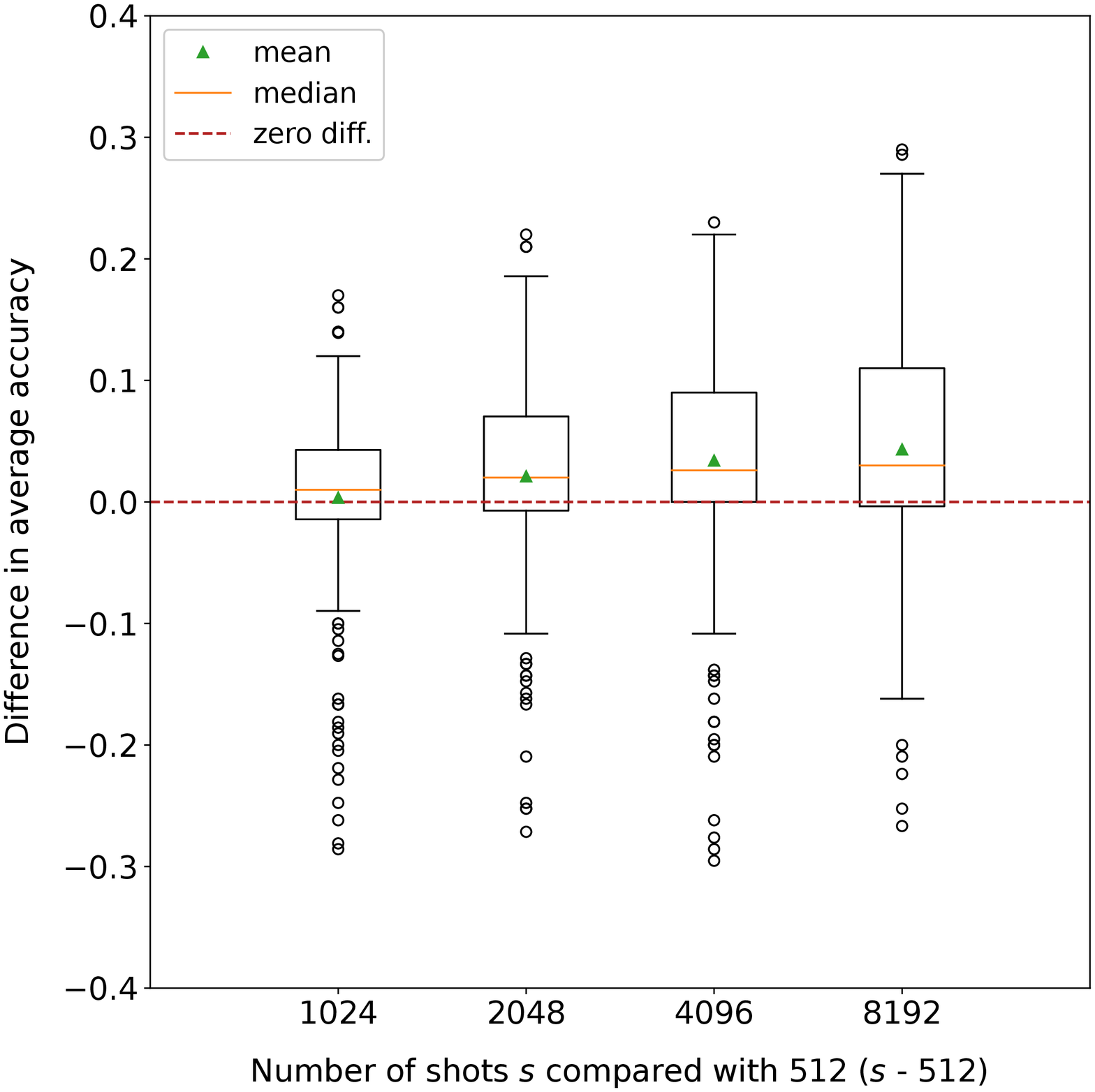}
    } 
    \subfloat[\label{fig:shots-num-comp-ji}]{
        \includegraphics[width=0.48\linewidth]{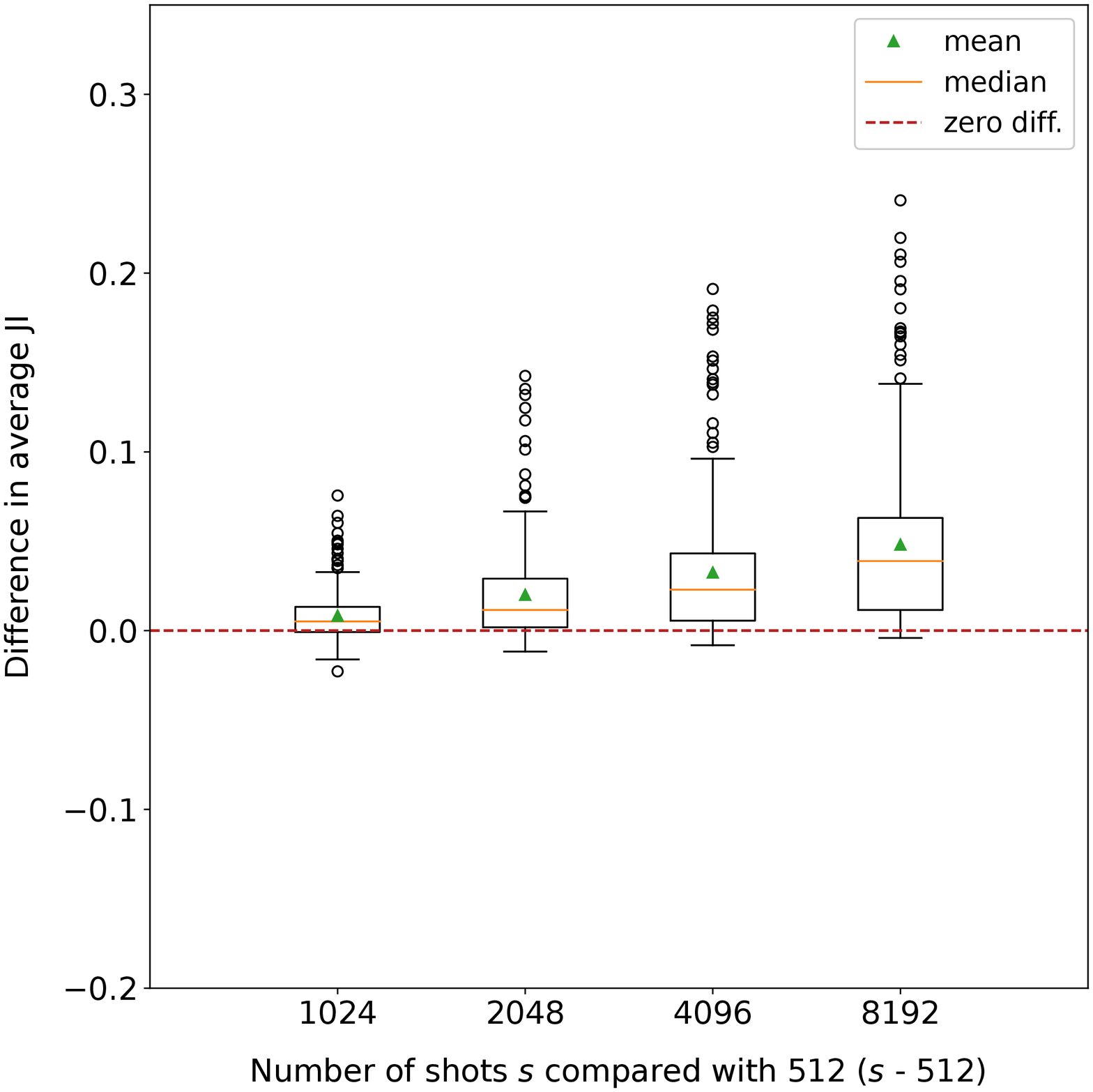}
    }
    \caption{Comparison of different numbers of shots in terms of accuracy (a) and Jaccard index (b) for the \textit{simulation} execution modality in the (\textit{extension}, \textit{avg}) configuration. Each data point corresponds to the difference for a (\textit{dataset fold}, \textit{k value}) pair}
    \label{fig:shots-num-comp}

    \vspace{10pt}

    \hypersetup{hidelinks}
    \centering
    \captionof{table}{Wilcoxon signed-rank test and one-sample T-test applied to the distributions shown in \cref{fig:shots-num-comp-acc} (a) and \cref{fig:shots-num-comp-ji} (b). The values reported in the tables are the p-values obtained ($\alpha\,{=}\,0.05$)}
    \label{tab:shots-num-comp-stats}
    \small
    \captionsetup{position=top}
    \subfloat[\label{tab:shots-num-comp-acc-stats}]{
        \begin{tabular}{c|c|c|c|c}
                     & \textbf{1024-512} & \textbf{2048-512} & \textbf{4096-512} & \textbf{8192-512} \\ \hline
    		Wilcoxon & 0.001             & 3.131E-07         & 2.432E-11         & 6.830E-11         \\ \hline
    		T-test   & 0.473             & 5.648E-05         & 3.016E-08         & 7.904E-11         \\
        \end{tabular}
    } \\
    \subfloat[\label{tab:shots-num-comp-ji-stats}]{
        \begin{tabular}{c|c|c|c|c}
                     & \textbf{1024-512} & \textbf{2048-512} & \textbf{4096-512} & \textbf{8192-512} \\ \hline
    		Wilcoxon & 1.151E-17         & 5.672E-33         & 4.761E-38         & 1.393E-40         \\ \hline
    		T-test   & 4.330E-17         & 5.712E-26         & 1.955E-30         & 1.561E-38         \\ 
        \end{tabular}
    }
\end{figure}

\subsubsection{Number of shots analysis}
\label{subsubsec:num-shots-analysis}
The last analysis is devoted to the relationship between number of shots (measurements) and performance for the \textit{simulation} execution modality. In particular, for this investigation, only the best quantum $k$-NN configuration has been considered. Since the primary goal of the quantum $k$-NN is to correctly find the $k$ nearest neighbors, the (\textit{extension}, \textit{avg}) configuration has been used. Indeed, it has achieved the best Jaccard index and Average Jaccard score, as shown in \cref{subsubsec:encodings-and-dist-ests-comp}. The results are presented in \cref{fig:shots-num-comp} and \cref{fig:shots-num-comp-aj} (the latter is available in \cref{subsec:app-num-shots-analysis}) by means of difference boxplots in which 512 is employed as the baseline number of shots.

In practice, the performance tend to improve by increasing the number of shots, and the trend is more evident for both the Jaccard index (\cref{fig:shots-num-comp-ji}) and the Average Jaccard score (\cref{fig:shots-num-comp-aj}), although the differences in absolute value are smaller when compared to the ones for the accuracy (\cref{fig:shots-num-comp-acc}). Furthermore, almost all performance differences are statistically significant in terms of both median (Wilcoxon signed-rank test) and mean (one-sample T-test), as reported in \cref{tab:shots-num-comp-stats} and \cref{tab:shots-num-comp-aj-stats} (available in \cref{subsec:app-num-shots-analysis}); the only exception is represented by the $1024 - 512$ comparison in terms of mean for the accuracy. Eventually, it is worth highlighting that, the larger the dataset, the higher the number of shots required to estimate the joint probability values.

\section{Conclusion}
\label{sec:conclusion}
In this article, a novel quantum $k$-NN algorithm based on the Euclidean distance metric has been introduced. In detail, two new encodings of the input data into the quantum states amplitudes, with different properties and low qubit requirements, have been presented (these encodings do not require the unit-norm normalization of the input data). The quantum circuit employed, which does not involve oracles, performs a SWAP-test-like procedure characterised by a fixed number of elementary gates; in this way, quantities related to the pairwise Euclidean distances are computed in parallel. Eventually, given the measurements results (the measurements must be repeated several times), two different ways of estimating the Euclidean distance values have been illustrated; the final training data sorting and classification are classical. 

In addition to the theoretical formulation and some complexity observations, an implementation of the algorithm in Python and an extensive empirical evaluation have been provided. First of all, the experimental results have confirmed the correctness of the formulation, with the \textit{statevector} execution modality (ideal execution with an infinite number of shots) achieving the same performance as the \textit{classical} one; it is worth remarking that the advantage over the classical counterpart lies in the execution time. As expected, \textit{statevector} has outperformed \textit{simulation}, for which the number of measurements is limited, in both classification accuracy and correctness of the nearest neighbors found (the difference is more marked for the latter). Among the (\textit{encoding}, \textit{distance estimate}) configurations tested, (\textit{translation}, \textit{avg}) has achieved the best results in terms of classification accuracy, whereas (\textit{extension}, \textit{avg}) has found the best nearest neighbors. It is worth highlighting that, in the two configurations just mentioned, the encoding is different, while the distance estimate is the same; however, since the primary goal of the algorithm is to find the correct nearest neighbors, (\textit{extension}, \textit{avg}) can be considered the best configuration overall. Concerning the classical baseline methods considered, half of them have achieved better results than the quantum $k$-NN in the \textit{statevector} modality, whereas the quantum $k$-NN in the \textit{simulation} modality has been always outperformed. Eventually, the analysis on the number of shots has certified that the performance of the algorithm in \textit{simulation} can be improved by increasing the number of measurements.

Possible future work includes testing the model presented here on different datasets, with a higher number of shots, and on real quantum machines, which are characterised by the presence of noise.

\backmatter


\section*{Statements and Declarations}

\textbf{Funding} This work was supported by Q@TN, the joint lab between University of Trento, FBK-Fondazione Bruno Kessler, INFN-National Institute for Nuclear Physics and CNR-National Research Council. In addition, this work was partially supported by project SERICS (PE00000014) under the MUR National Recovery and Resilience Plan funded by the European Union - NextGenerationEU. Eventually, the authors gratefully acknowledge the Italian Ministry of University and Research (MUR), which, under the initiative "Dipartimenti di Eccellenza 2018-2022 (Legge 232/2016)", has provided the computational resources used in the experiments. \\

\noindent\textbf{Competing interests} The authors declare that no competing interests exist. \\

\noindent\textbf{Ethics approval} Not applicable. \\

\noindent\textbf{Consent to participate} Not applicable. \\

\noindent\textbf{Consent for publication} Not applicable. \\

\noindent\textbf{Availability of data and materials} The data that support the findings of this study (datasets, information needed to reproduce the experiments, and collected results data) are available in the Figshare repository, \url{https://doi.org/10.6084/m9.figshare.22598455.v1}. In addition, the same data, together with the raw results data, are reachable from the GitHub repository, \url{https://github.com/ZarHenry96/euclidean-quantum-k-nn}. \\

\noindent\textbf{Code availability} The code is available in the GitHub repository, \url{https://github.com/ZarHenry96/euclidean-quantum-k-nn}. \\

\noindent\textbf{Authors' contributions} Conceptualization: Enrico Zardini, Enrico Blanzieri, Davide Pastorello; Data curation: Enrico Zardini; Formal Analysis: Enrico Zardini, Enrico Blanzieri, Davide Pastorello; Investigation: Enrico Zardini; Methodology: Enrico Zardini, Enrico Blanzieri, Davide Pastorello; Software: Enrico Zardini; Supervision: Enrico Blanzieri, Davide Pastorello; Validation: Enrico Zardini; Visualization: Enrico Zardini; Original draft preparation: Enrico Zardini; Review and editing: Enrico Zardini, Enrico Blanzieri, Davide Pastorello.

\newcounter{mainFigure}
\setcounter{mainFigure}{\value{figure}}
\newcounter{mainTable}
\setcounter{mainTable}{\value{table}}
\newcounter{mainEquation}
\setcounter{mainEquation}{\value{equation}}

\begin{appendices}

\renewcommand{\thefigure}{\arabic{figure}}
\renewcommand{\theHfigure}{\arabic{figure}}
\renewcommand{\thetable}{\arabic{table}}
\renewcommand{\theHtable}{\arabic{table}}
\renewcommand{\theequation}{\arabic{equation}}
\renewcommand{\theHequation}{\arabic{equation}}
\setcounter{figure}{\value{mainFigure}}
\setcounter{table}{\value{mainTable}}
\setcounter{equation}{\value{mainEquation}}

\crefalias{section}{appendix}
\crefalias{subsection}{appendix}

\section{Derivations}
\label{sec:app-derivations}
Some derivations related to the output state $\ket{\gamma}$ (Equation \ref{eq:output-gamma-state}) are provided in this appendix.

\subsection{Probability of measuring 1 on the first qubit}
\label{subsec:app-state-one-prob}
The probability of measuring $1$ on the first qubit of the state $\ket{\gamma}$ is
\begin{align*}
    P(1) &= \left\| \ket{1}\bra{1} \ket{\gamma} \right\|^2 = \\
         &= \left\| \frac{1}{2} \ket{1} \otimes \left(\frac{1}{\sqrt{2}}(\ket{0}\ket{\alpha} - \ket{0}\ket{\beta} + \ket{1}\ket{\beta} - \ket{1}\ket{\alpha})\right) \right\|^2 = \\
         &= \!\begin{multlined}[t][10.5cm]\frac{1}{8}(\bra{0}\bra{\alpha} - \bra{0}\bra{\beta} + \bra{1}\bra{\beta} - \bra{1}\bra{\alpha}) \times \\ \times (\ket{0}\ket{\alpha} - \ket{0}\ket{\beta} + \ket{1}\ket{\beta} - \ket{1}\ket{\alpha}) = \end{multlined} \\
         &= \frac{1}{8} (1 - \bra{\alpha}\ket{\beta} - \bra{\beta}\ket{\alpha} + 1 + 1 - \bra{\beta}\ket{\alpha} - \bra{\alpha}\ket{\beta} + 1) = \\
         &= \frac{1}{8} (4 - 2 \bra{\alpha}\ket{\beta} - 2 \bra{\beta}\ket{\alpha}) = \tag{$\ket{\alpha}$ and $\ket{\beta}$ have real coefficients}\\
         &= \frac{1}{8} (4 - 4 \bra{\alpha}\ket{\beta}) = \\
         & = \frac{1}{2} (1 - \bra{\alpha}\ket{\beta}).
\end{align*}

\subsection{Reduced final state}
\label{subsec:app-reduced-final-state}
Given $\ket{\gamma}$, let us first pull out the summation on the index register $\ket{j}$ inside $\ket{\alpha}$ and $\ket{\beta}$. In this way, we obtain a state in the form
\begin{equation*}
    \frac{1}{\sqrt{N}} \sum_{j=0}^{N-1}\left[...\right]\ket{j},
\end{equation*}
where $\left[...\right]$ includes all circuit qubits except those belonging to the index register. Then, let us trace out, namely, discard, the second qubit in the circuit and the features register $\ket{i}$; from the mathematical viewpoint, this corresponds to compute the partial trace over these qubits of the density operator describing the system. By doing this, we obtain a reduced version of the final state including only the first qubit in the circuit and the index register, which can be written as
\begin{equation*}
    \frac{1}{\sqrt{N}} \sum_{j=0}^{N-1}\left[\sqrt{P(0\mid j)}\ket{0} + \sqrt{P(1\mid j)}\ket{1}\right]\ket{j}.
\end{equation*}
Eventually, let us exploit the derivations presented in \cref{subsec:app-state-one-prob}. In practice, as shown in \cref{eq:s-value}, $P(1\mid j)$ turns out to be equal to $\frac{1}{2}(1 - \langle \mathvec{x}_j, \mathvec{x}'_j\rangle)$, since the summation on the index register (together with its coefficient) has been pulled out. In addition, $P(0\mid j)$ must be equal to $1 - P(1\mid j)$ due to the law of total probability. This leads to the definition of the reduced final state $\ket{\delta}$ provided in \cref{eq:final-delta-state}.

\section{Distance estimates}
\label{sec:app-dist-estimates}
In this appendix, some additional information about the \textit{avg} and \textit{diff} distance estimates is provided. In particular, let us consider two preprocessed training instances $\mathvec{v}_{j_1}$ and $\mathvec{v}_{j_2}$, with $j_1, j_2 \in \{0, ..., N-1\}$, and a preprocessed test instance $\mathvec{v}'$. Let $d_0(\mathvec{v}_{j_1}, \mathvec{v}')$ and $d_1(\mathvec{v}_{j_1}, \mathvec{v}')$ be the Euclidean distances from $\mathvec{v}'$ estimated from the joint probabilities $P(0, j_1)$ and $P(1, j_1)$ for $\mathvec{v}_{j_1}$ (analogously for $\mathvec{v}_{j_2}$). Then, the following relationships hold:
\begin{align*}
    \mathit{avg} &= \frac{d_0(\mathvec{v}_{j_1}, \mathvec{v}') + d_1(\mathvec{v}_{j_1}, \mathvec{v}')}{2}, \\ 
    \mathit{diff} &= \sqrt{\frac{d_0(\mathvec{v}_{j_1}, \mathvec{v}')^2 + d_1(\mathvec{v}_{j_1}, \mathvec{v}')^2}{2}}. 
\end{align*}
The former is just the definition of the \textit{avg} distance estimate, whereas the latter can be easily verified using \cref{eq:extension-eucl-dist} (or \ref{eq:translation-eucl-dist}, depending on the encoding selected) together with \cref{eq:joint-p0,eq:joint-p1}.

\subsection{Instance sorting}
\label{subsec:instance-sorting}
It is possible that $\mathvec{v}_{j_1}$ and $\mathvec{v}_{j_2}$ are sorted differently according to the \textit{avg} and \textit{diff} distance estimates. Indeed, let us consider the following scenario:
\begin{align*}
    d_0(\mathvec{v}_{j_1}, \mathvec{v}') &= 0.5   &  d_0(\mathvec{v}_{j_2}, \mathvec{v}') &= 0.4  \\
    d_1(\mathvec{v}_{j_1}, \mathvec{v}') &= 0.29  &  d_1(\mathvec{v}_{j_2}, \mathvec{v}') &= 0.4  \,.  
\end{align*}
In this case, the \textit{avg} distance estimates for $\mathvec{v}_{j_1}$ and $\mathvec{v}_{j_2}$ are $0.395$ and $0.4$, respectively, while the \textit{diff} distance estimates are $0.409$ and $0.4$, respectively. Hence, $\mathvec{v}_{j_1}$ turns out to be closer than $\mathvec{v}_{j_2}$ (to $\mathvec{v}'$) according to \textit{avg} and further than it according to \textit{diff}.

\subsection{Magnitude}
\label{subsec:magnitude}
Let us assume that $d_0(\mathvec{v}_{j_1}, \mathvec{v}')$ and $d_1(\mathvec{v}_{j_1}, \mathvec{v}')$ can be mathematically computed, namely, the arguments of the square root in \cref{eq:extension-eucl-dist} (or \ref{eq:translation-eucl-dist}, depending on the encoding selected) are non-negative. Then, the \textit{avg} distance estimate is always lower than or equal to the corresponding \textit{diff} estimate. In fact, the opposite would be true if and only if $(d_0(\mathvec{v}_{j_1}, \mathvec{v}') - d_1(\mathvec{v}_{j_1}, \mathvec{v}'))^2 < 0$, which is not possible.

If the assumption about the square root arguments does not apply due to the state count distribution obtained, the \textit{avg} distance estimate might turn out to be higher than the corresponding \textit{diff} estimate. Indeed, in the implementation provided here, square roots of negative values are approximated to zero, as explained in \cref{sec:implementation}.

\section{Additional plots}
\label{sec:app-additional-plots}
Additional results plots and related statistical significance tables are provided in this appendix.
\clearpage

\subsection{Execution modalities comparison}
\label{subsec:app-exec-modalities-comp}

\begin{figure}[h!]
    \centering
    \subfloat[\label{fig:sv-vs-si-ext-diff-acc}]{
        \centering
        \includegraphics[width=0.31\linewidth]{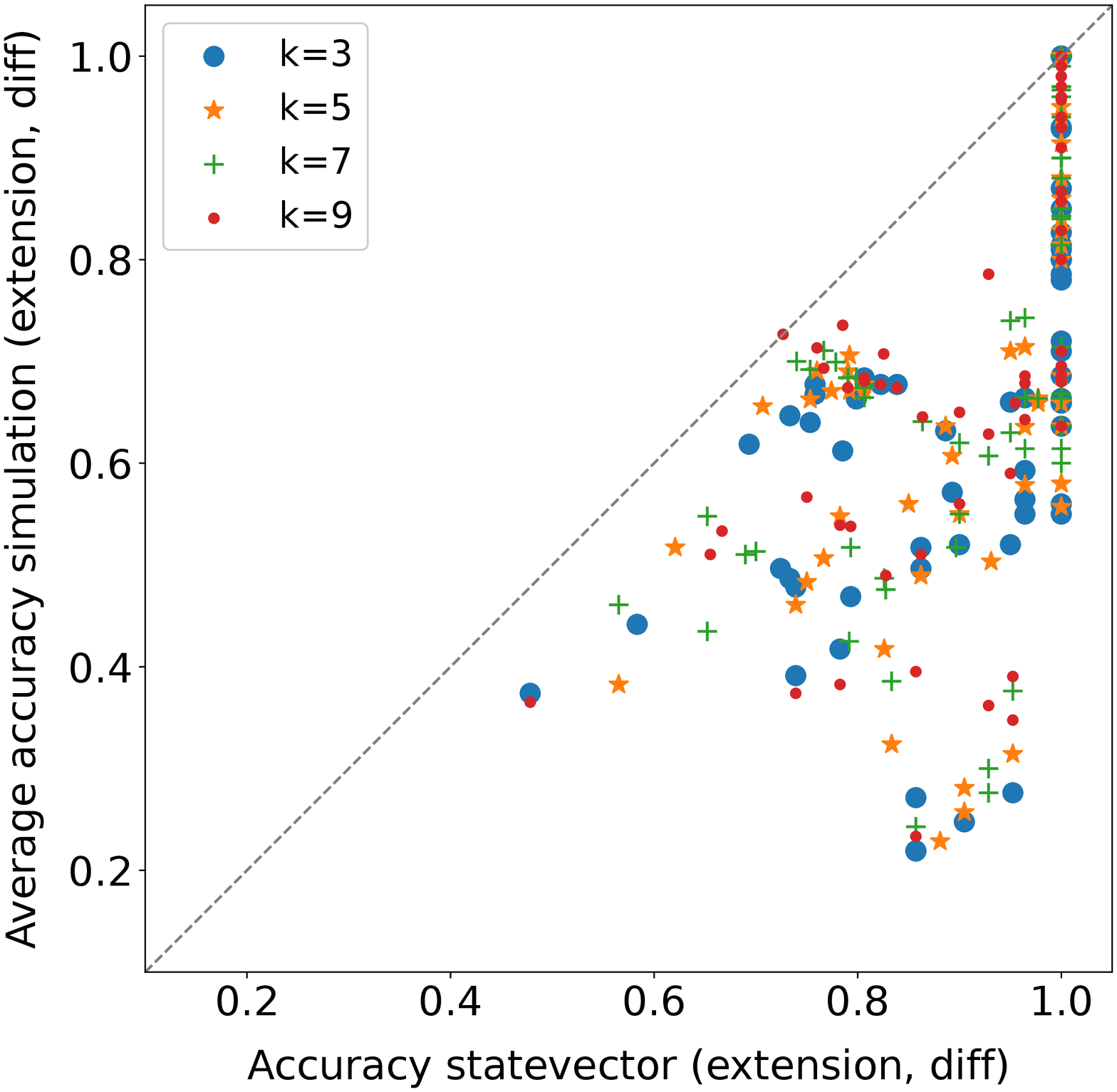}
    }
    \subfloat[\label{fig:sv-vs-si-ext-diff-ji}]{
        \centering
        \includegraphics[width=0.31\linewidth]{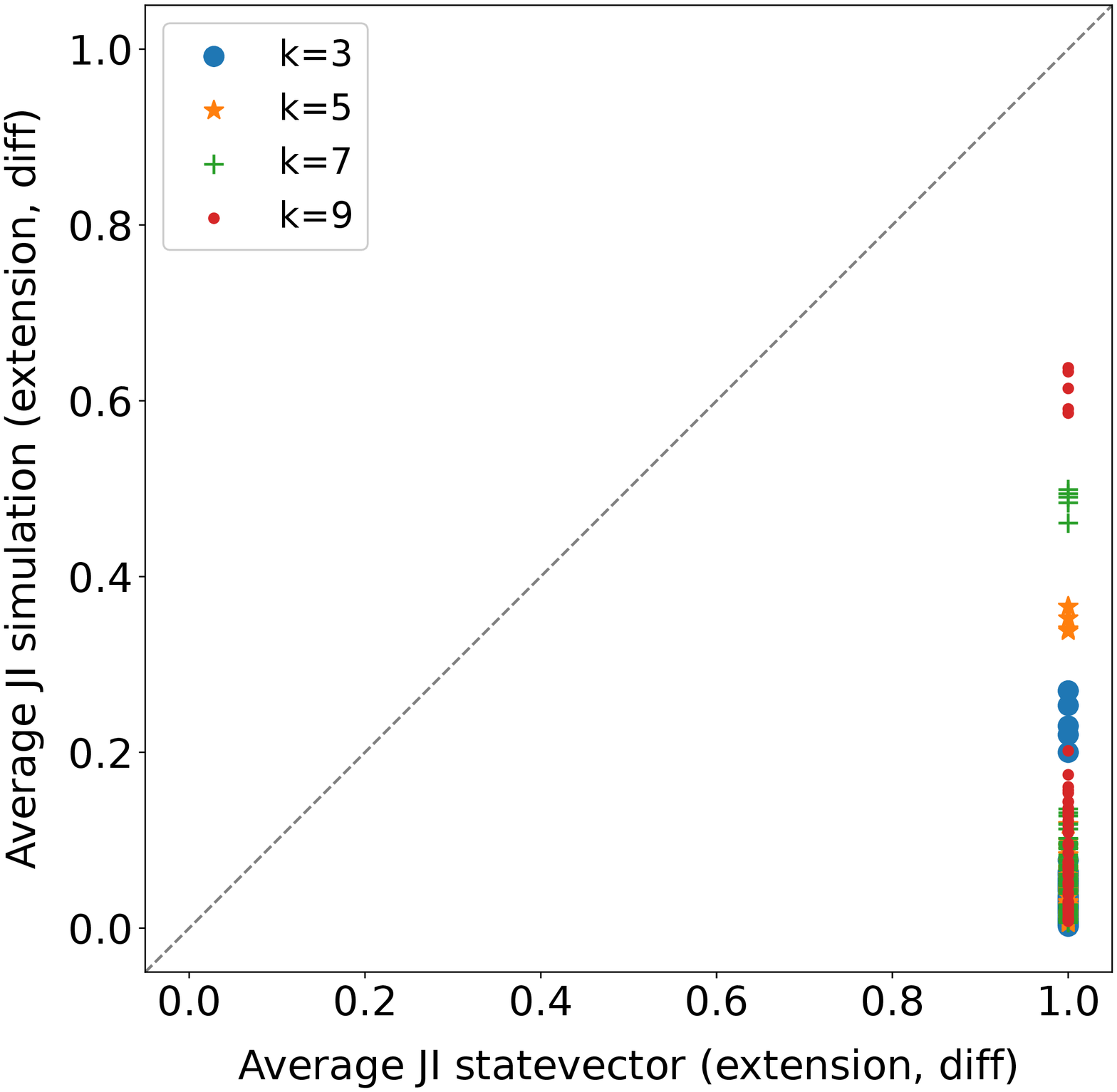}
    }
    \subfloat[\label{fig:sv-vs-si-ext-diff-aj}]{
        \centering
        \includegraphics[width=0.31\linewidth]{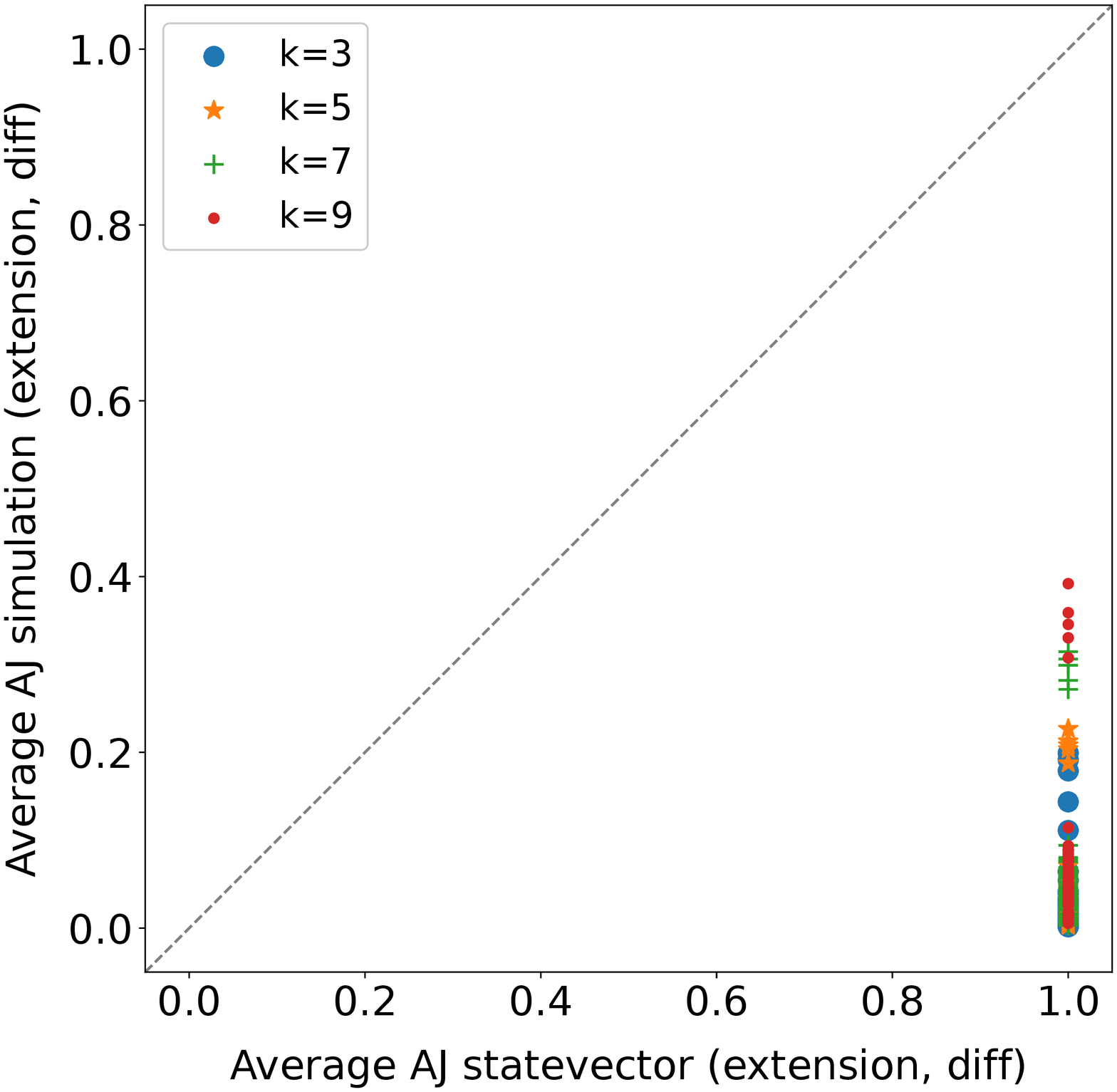}
    }
    \caption{Comparison between \textit{statevector} (\textit{extension}, \textit{diff}) and \textit{simulation} (\textit{extension}, \textit{diff}) in terms of accuracy (a), Jaccard index (b), and Average Jaccard score (c). The number of shots for \textit{simulation} is 1024, and each point is related to a dataset fold}
    \label{fig:sv-vs-si-ext-diff}

    \vspace{10pt}

    \hypersetup{hidelinks}
    \centering
    \captionof{table}{Wilcoxon signed-rank test ($\alpha\,{=}\,0.05$) applied to the distributions shown in \cref{fig:sv-vs-si-ext-diff}. The values reported in the table are the p-values obtained}
    \label{tab:sv-vs-si-ext-diff-stats}
    \small
    \begin{tabular}{c|c|c|c|c}
                                         & \textbf{k=3} & \textbf{k=5} & \textbf{k=7} & \textbf{k=9} \\ \hline
		\cref{fig:sv-vs-si-ext-diff-acc} & 1.106E-10    & 1.106E-10    & 7.530E-11    & 1.719E-10    \\ \hline
		\cref{fig:sv-vs-si-ext-diff-ji}  & 1.628E-11    & 1.629E-11    & 1.630E-11    & 1.630E-11    \\ \hline
		\cref{fig:sv-vs-si-ext-diff-aj}  & 1.630E-11    & 1.630E-11    & 1.630E-11    & 1.630E-11    \\
    \end{tabular}
\end{figure}

\begin{figure}[h!]
    \centering
    \subfloat[\label{fig:sv-vs-si-transl-avg-acc}]{
        \centering
        \includegraphics[width=0.31\linewidth]{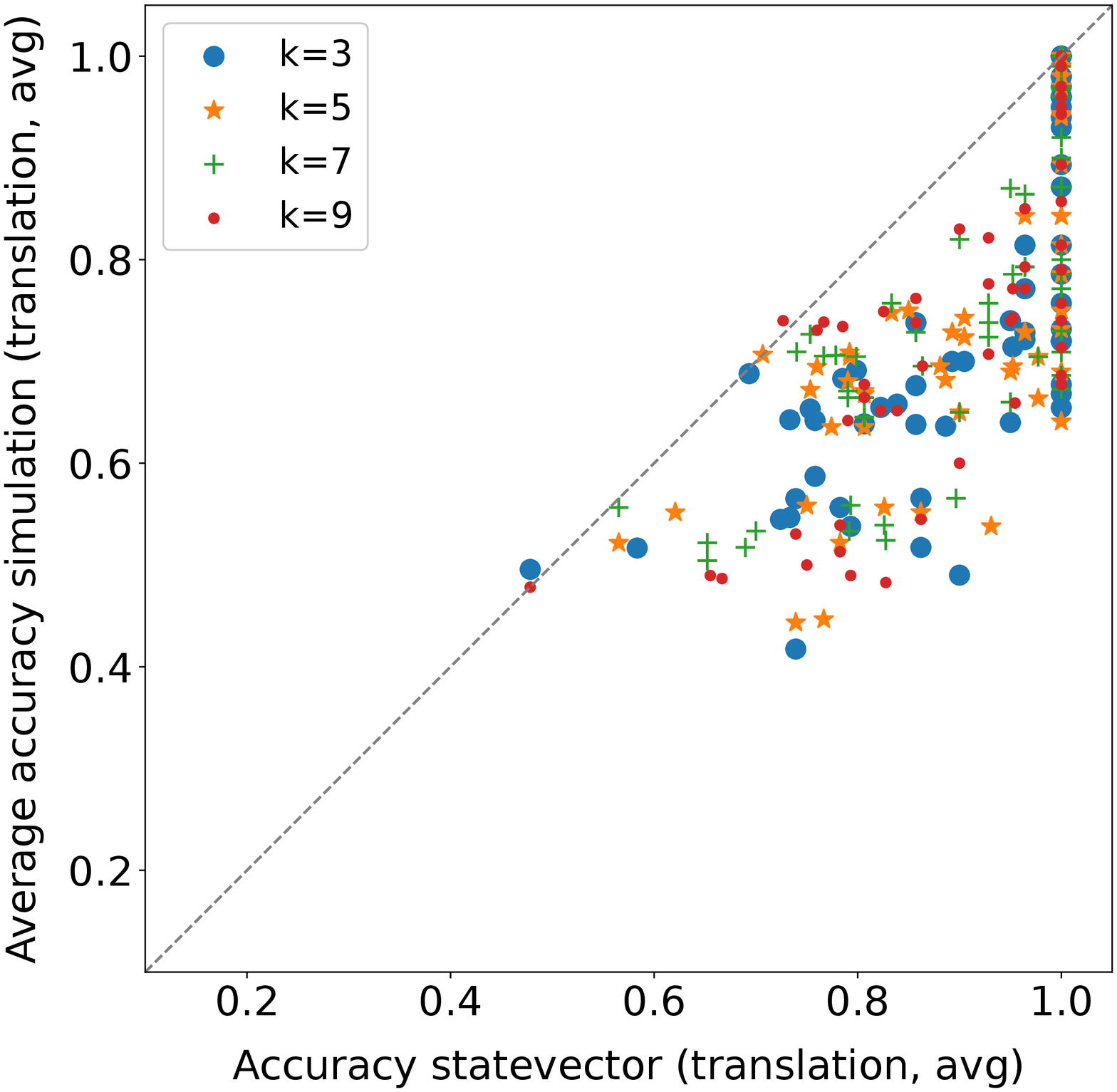}
    }
    \subfloat[\label{fig:sv-vs-si-transl-avg-ji}]{
        \centering
        \includegraphics[width=0.31\linewidth]{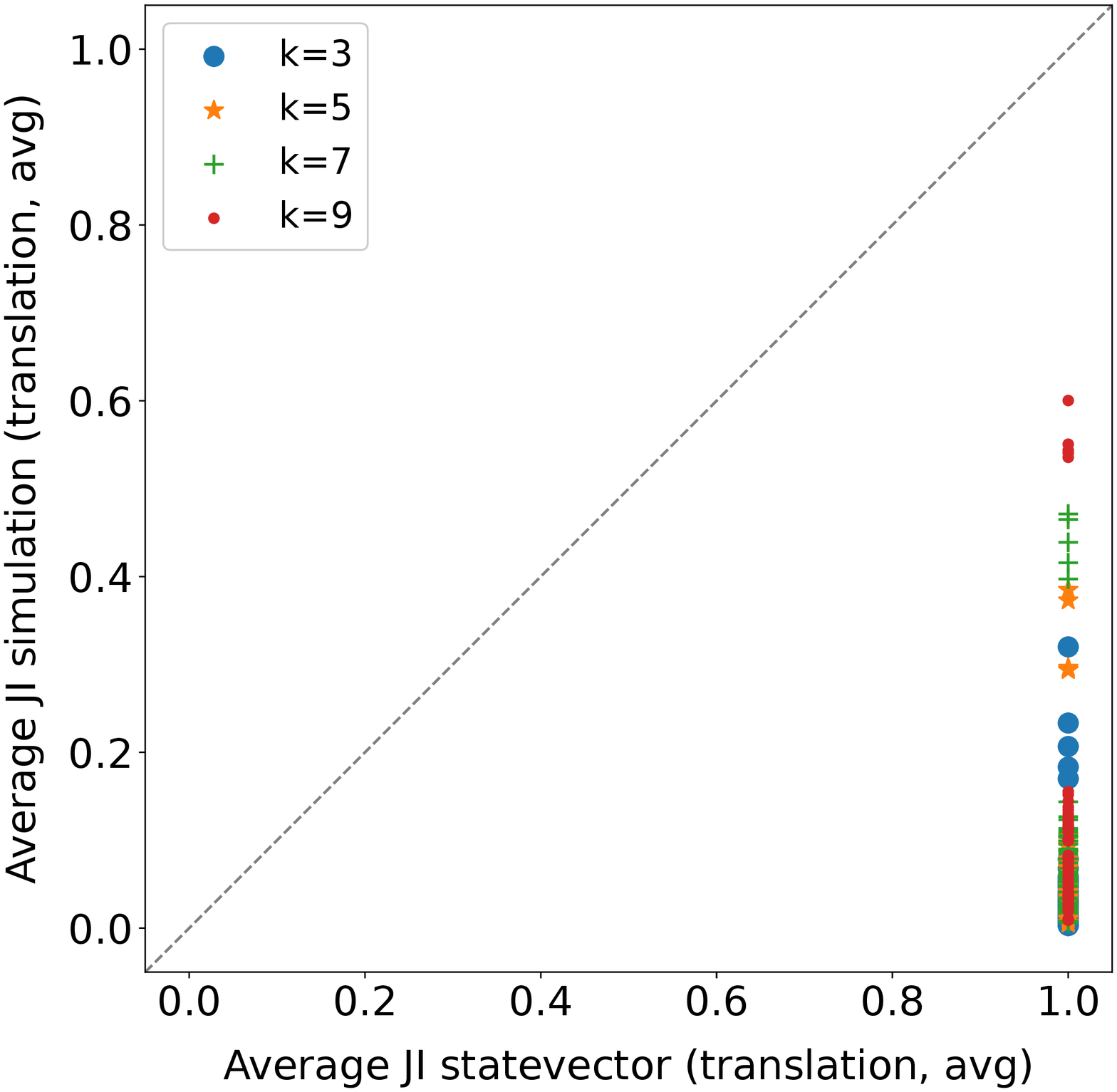}
    } 
    \subfloat[\label{fig:sv-vs-si-transl-avg-aj}]{
        \centering
        \includegraphics[width=0.31\linewidth]{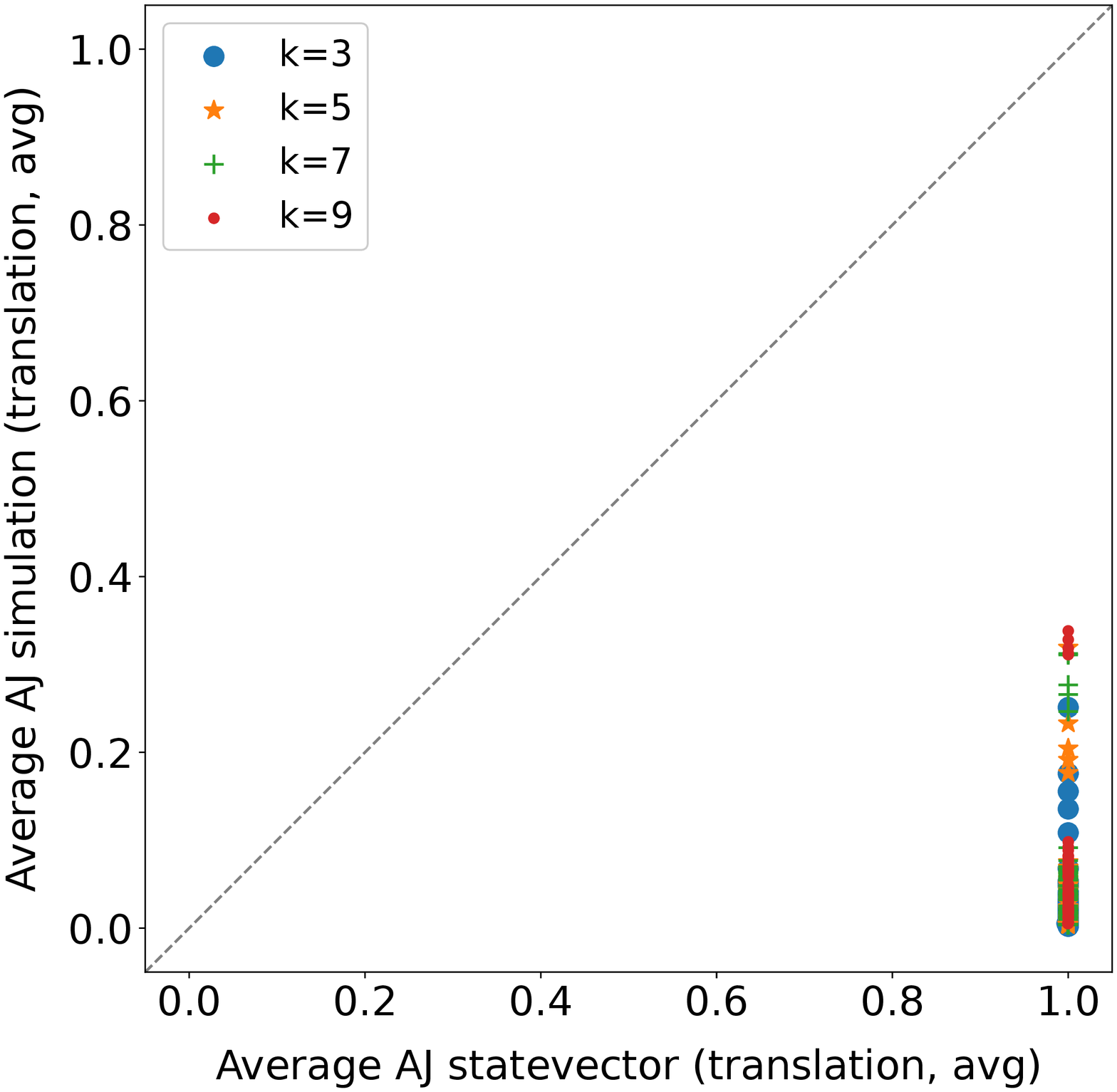}
    }
    \caption{Comparison between \textit{statevector} (\textit{translation}, \textit{avg}) and \textit{simulation} (\textit{translation}, \textit{avg}) in terms of accuracy (a), Jaccard index (b), and Average Jaccard score (c). The number of shots for \textit{simulation} is 1024, and each point is related to a dataset fold}
    \label{fig:sv-vs-si-transl-avg}

    \vspace{10pt}

    \hypersetup{hidelinks}
    \centering
    \captionof{table}{Wilcoxon signed-rank test ($\alpha\,{=}\,0.05$) applied to the distributions shown in \cref{fig:sv-vs-si-transl-avg}. The values reported in the table are the p-values obtained}
    \label{tab:sv-vs-si-transl-avg-stats}
    \small
    \begin{tabular}{c|c|c|c|c}
                                           & \textbf{k=3} & \textbf{k=5} & \textbf{k=7} & \textbf{k=9} \\ \hline
		\cref{fig:sv-vs-si-transl-avg-acc} & 1.234E-10    & 3.492E-10    & 3.492E-10    & 6.149E-10    \\ \hline
		\cref{fig:sv-vs-si-transl-avg-ji}  & 1.628E-11    & 1.630E-11    & 1.630E-11    & 1.630E-11    \\ \hline
		\cref{fig:sv-vs-si-transl-avg-aj}  & 1.630E-11    & 1.630E-11    & 1.630E-11    & 1.630E-11    \\
    \end{tabular}
\end{figure}

\clearpage

\begin{figure}[t!]
    \centering
    \subfloat[\label{fig:sv-vs-si-transl-diff-acc}]{
        \centering
        \includegraphics[width=0.31\linewidth]{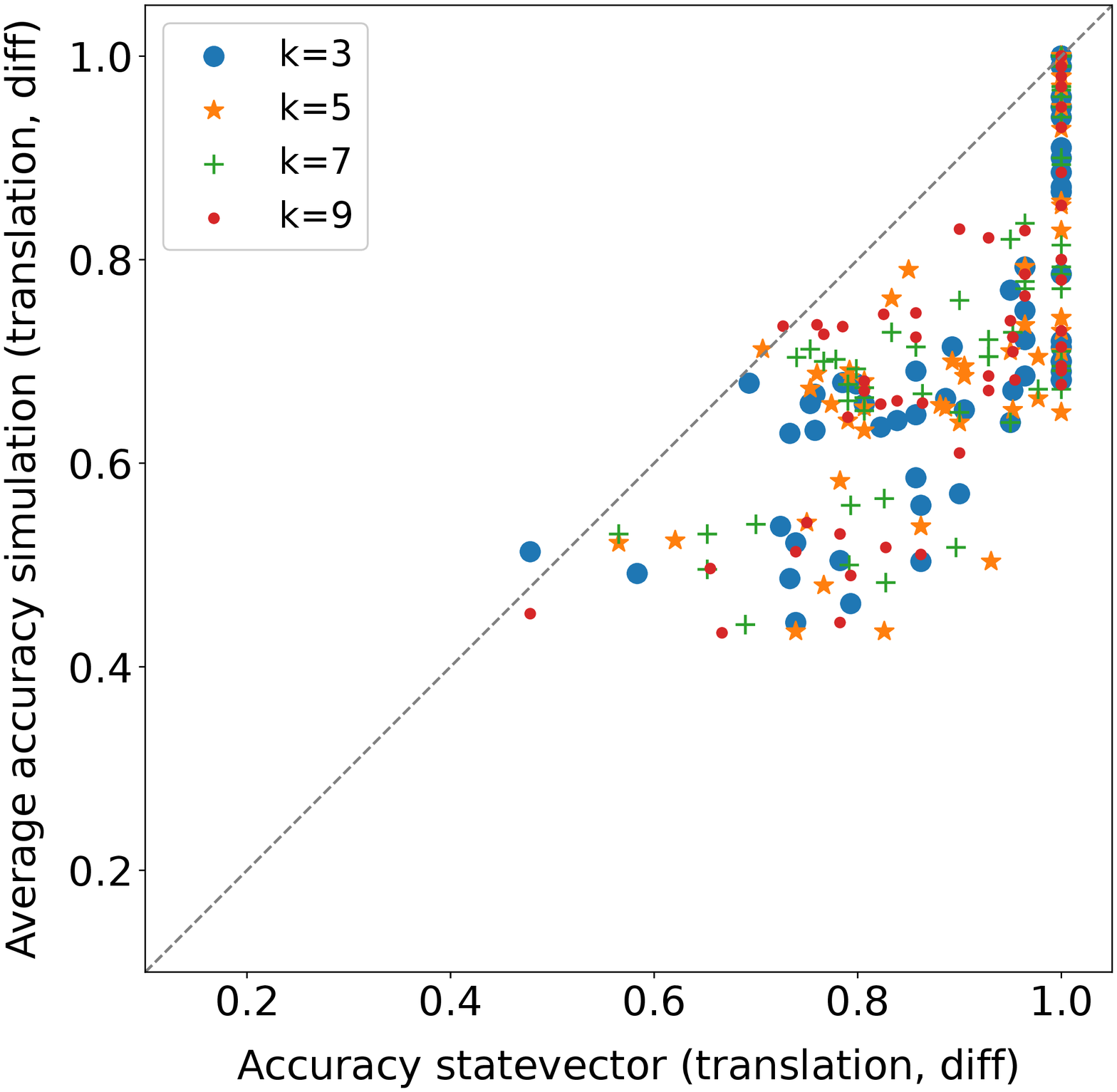}
    }
    \subfloat[\label{fig:sv-vs-si-transl-diff-ji}]{
        \centering
        \includegraphics[width=0.31\linewidth]{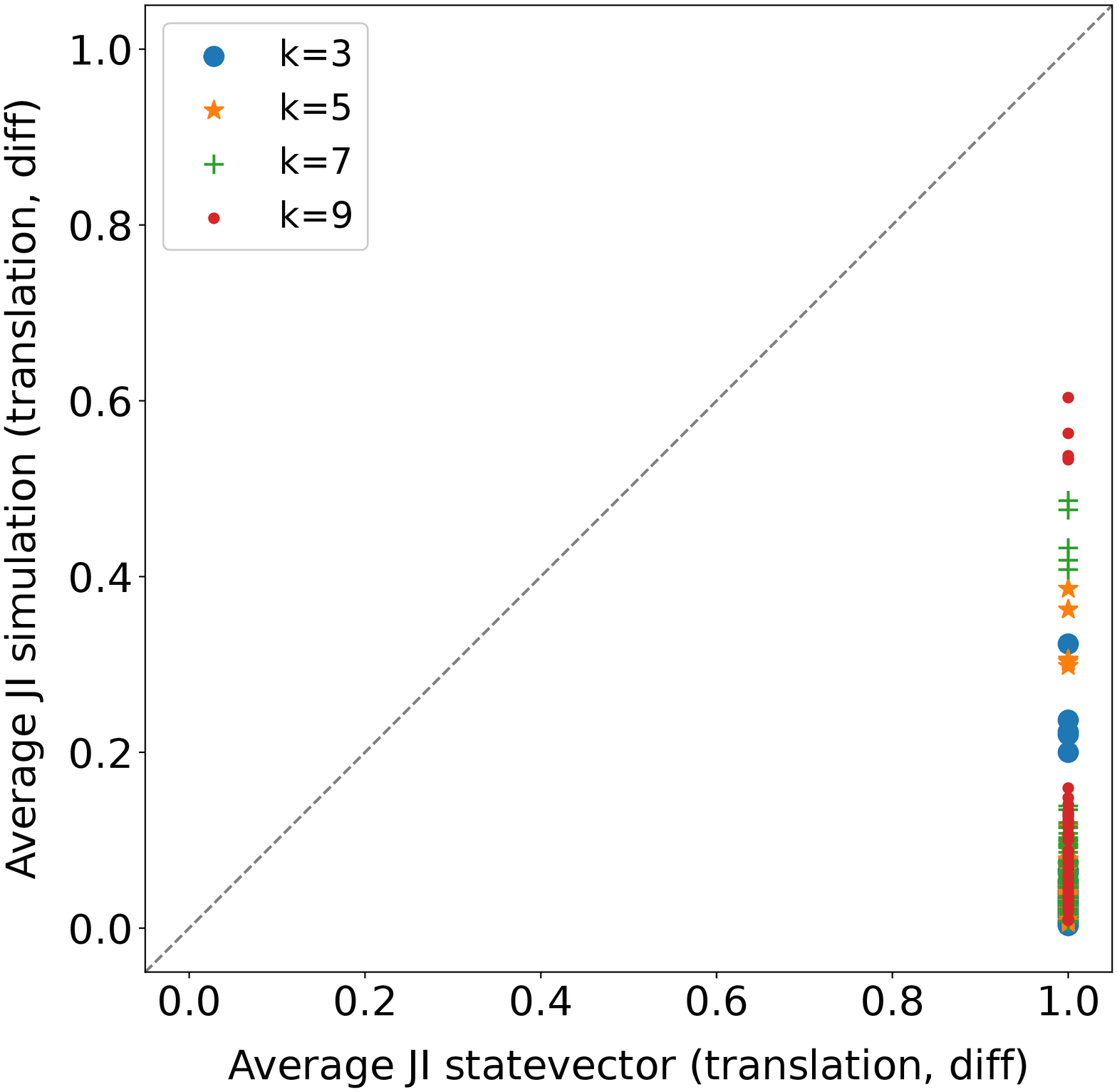}
    } 
    \subfloat[\label{fig:sv-vs-si-transl-diff-aj}]{
        \centering
        \includegraphics[width=0.31\linewidth]{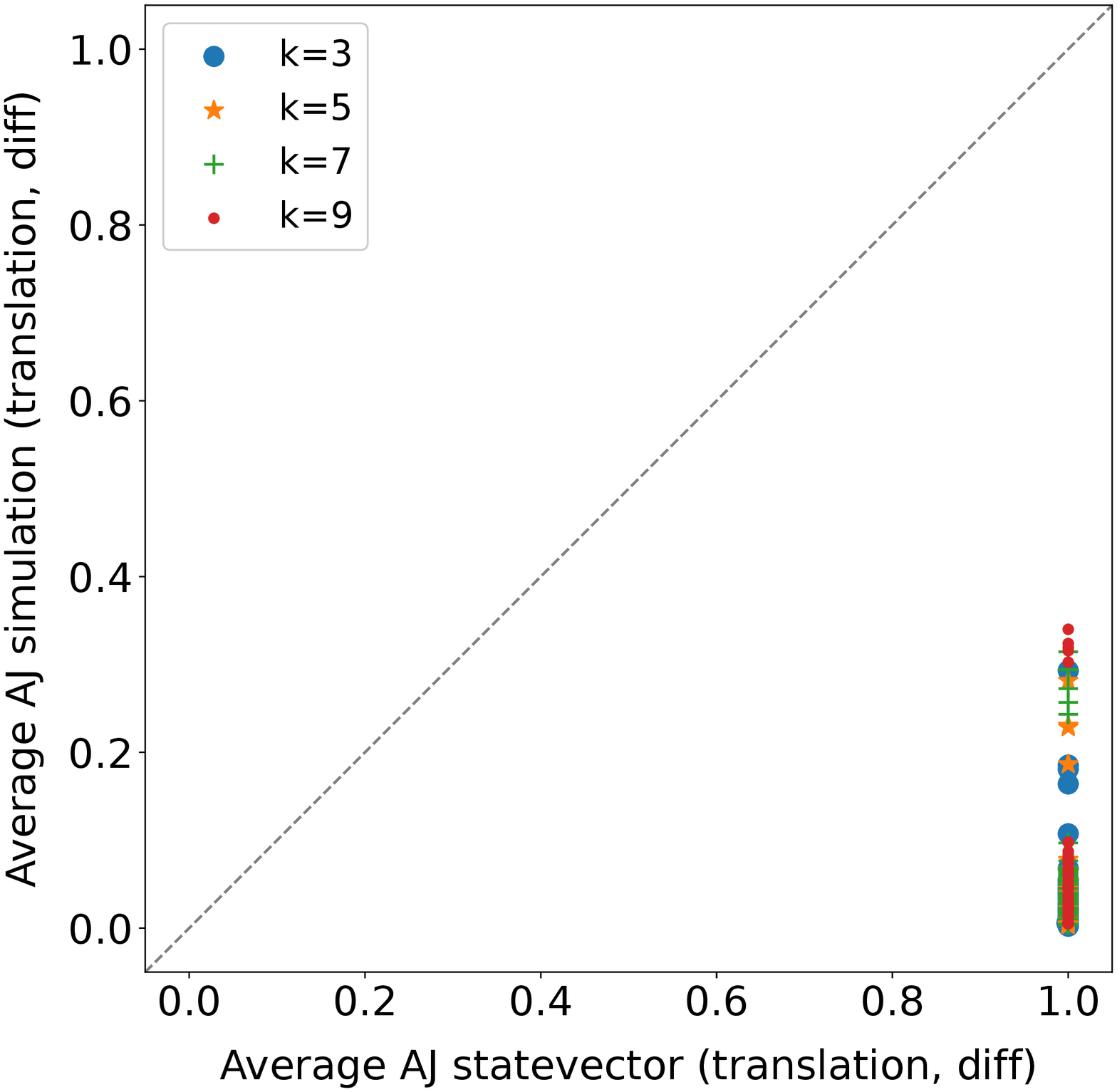}
    }
    \caption{Comparison between \textit{statevector} (\textit{translation}, \textit{diff}) and \textit{simulation} (\textit{translation}, \textit{diff}) in terms of accuracy (a), Jaccard index (b), and Average Jaccard score (c). The number of shots for \textit{simulation} is 1024, and each point is related to a dataset fold}
    \label{fig:sv-vs-si-transl-diff}

    \vspace{10pt}

    \hypersetup{hidelinks}
    \centering
    \captionof{table}{Wilcoxon signed-rank test ($\alpha\,{=}\,0.05$) applied to the distributions shown in \cref{fig:sv-vs-si-transl-diff}. The values reported in the table are the p-values obtained}
    \label{tab:sv-vs-si-transl-diff-stats}
    \small
    \begin{tabular}{c|c|c|c|c}
                                            & \textbf{k=3} & \textbf{k=5} & \textbf{k=7} & \textbf{k=9} \\ \hline
		\cref{fig:sv-vs-si-transl-diff-acc} & 1.379E-10    & 1.712E-10    & 7.516E-11    & 2.523E-10    \\ \hline
		\cref{fig:sv-vs-si-transl-diff-ji}  & 1.627E-11    & 1.629E-11    & 1.630E-11    & 1.630E-11    \\ \hline
		\cref{fig:sv-vs-si-transl-diff-aj}  & 1.629E-11    & 1.630E-11    & 1.630E-11    & 1.630E-11    \\
    \end{tabular}
\end{figure}

\subsection{Encodings and distance estimates comparison}
\label{subsec:app-encodings-and-dist-ests-comp}

\begin{figure}[h!]
    \centering    
    \includegraphics[width=0.98\linewidth]{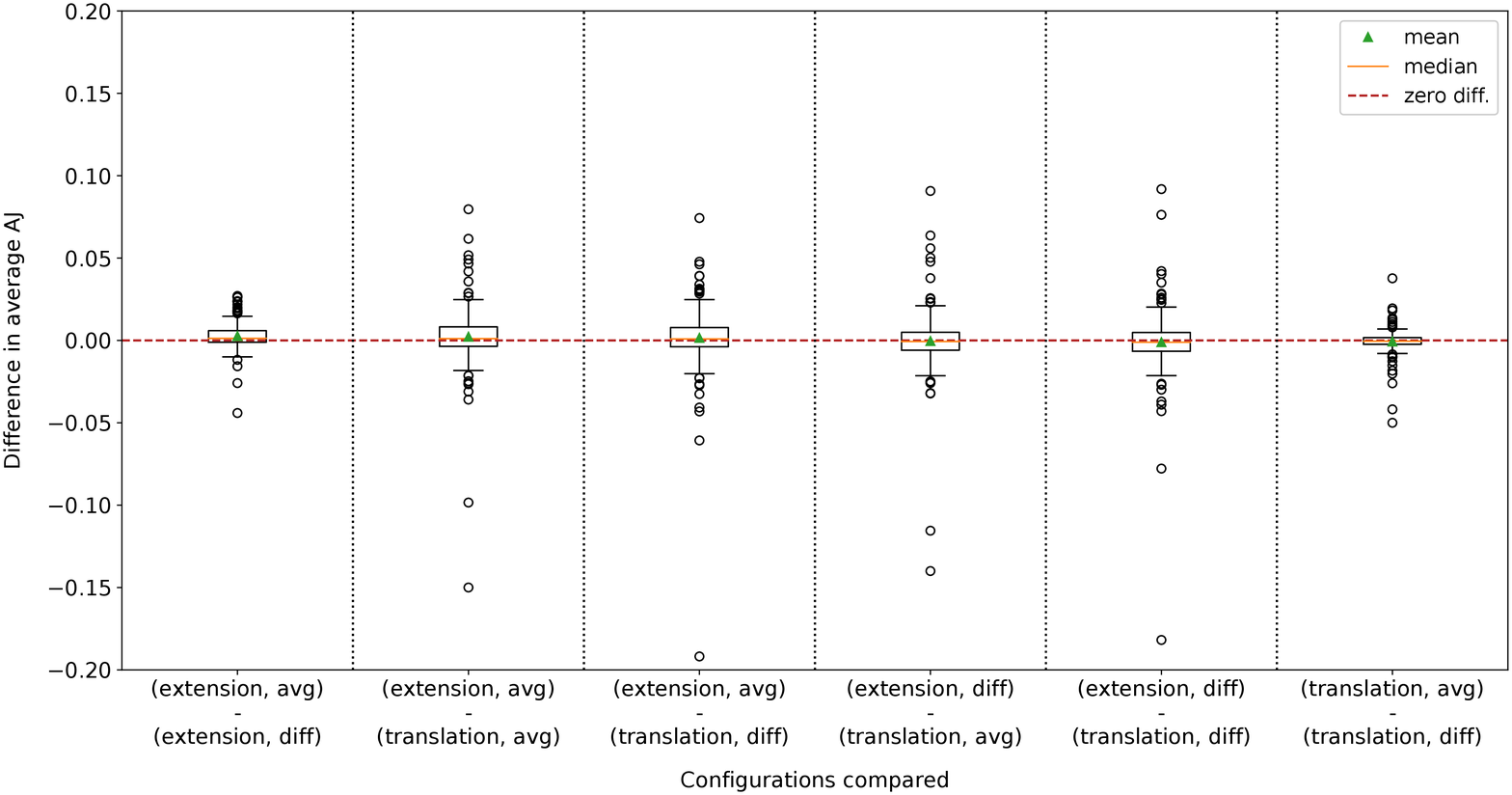}
    \caption{Comparison of (\textit{encoding}, \textit{distance estimate}) configurations in terms of Average Jaccard score for the \textit{simulation} execution modality. The number of shots is 1024, and each data point corresponds to the difference for a (\textit{dataset fold}, \textit{k value}) pair}
    \label{fig:encodings-and-dist-ests-comp-aj}
    
\end{figure}

\clearpage

\begin{table}[t]
    \hypersetup{hidelinks}
    \centering
    \caption{Wilcoxon signed-rank test and one-sample T-test applied to the distributions shown in \cref{fig:encodings-and-dist-ests-comp-aj}. Each column corresponds to a different comparison, and the first letter identifies the encoding (E=\textit{extension}, T=\textit{translation}), while the second letter identifies the distance estimate (A=\textit{avg}, D=\textit{diff}). The values reported in the table are the p-values obtained ($\alpha\,{=}\,0.05$)}
    \label{tab:encodings-and-dist-ests-comp-aj-stats}
    \begin{tabular}{c|c|c|c|c|c|c}
                 & \textbf{EA-ED} & \textbf{EA-TA} & \textbf{EA-TD} & \textbf{ED-TA} & \textbf{ED-TD} & \textbf{TA-TD} \\ \hline
        Wilcoxon & 1.556E-07      & 0.002          & 0.008          & 0.315          & 0.111          & 0.043          \\ \hline
        T-test   & 6.693E-07      & 0.057          & 0.202          & 0.735          & 0.374          & 0.123          \\
    \end{tabular}
\end{table}

\subsection{Comparison with baseline methods}
\label{subsec:app-comp-with-baselines}

\begin{figure}[h!]
    \centering
    \subfloat[\label{fig:si-transl-avg-vs-knn-cos}]{
        \centering
        \includegraphics[width=0.33\linewidth]{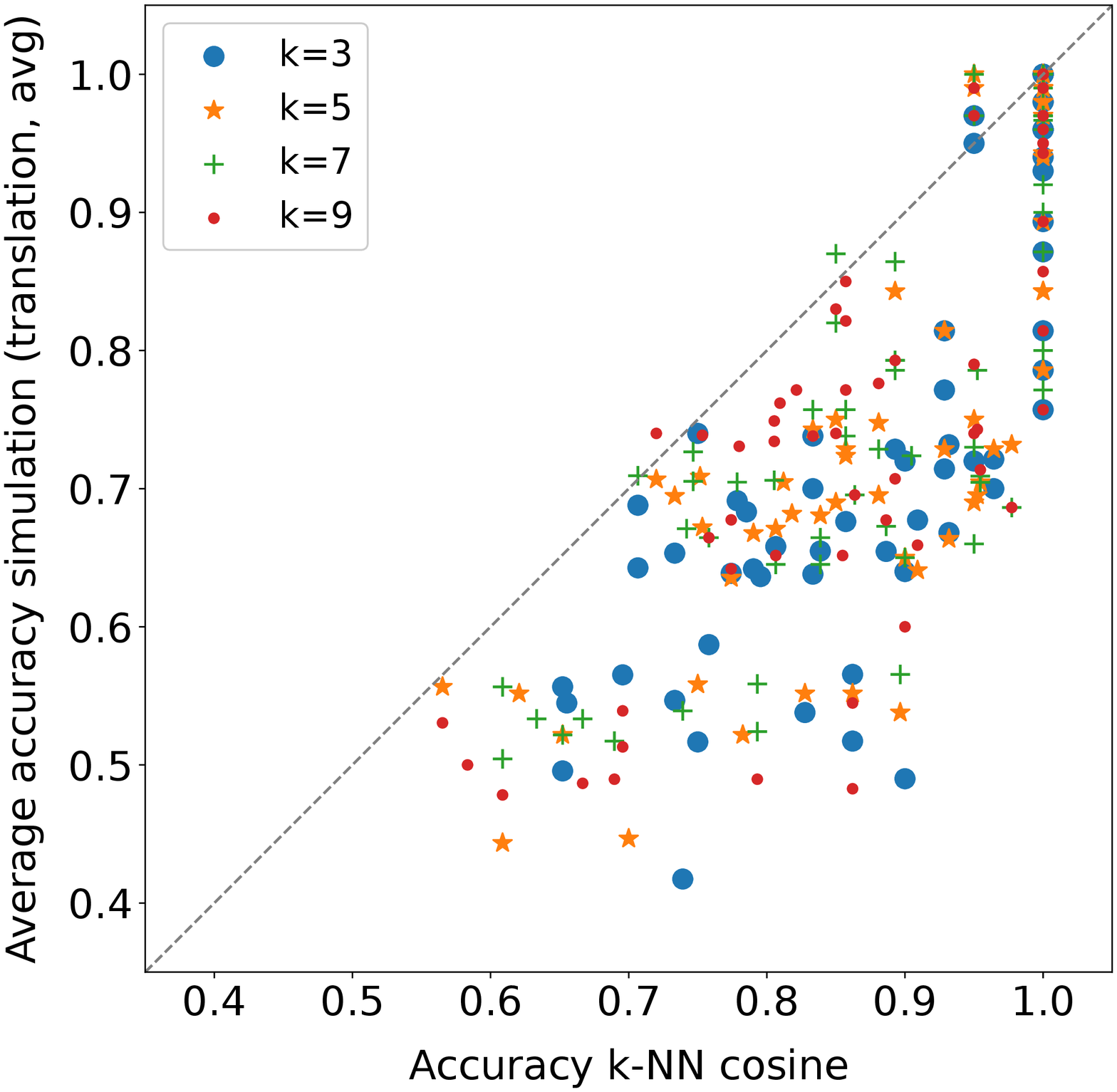}
    }
    \subfloat[\label{fig:si-transl-avg-vs-random-forest}]{
        \centering
        \includegraphics[width=0.33\linewidth]{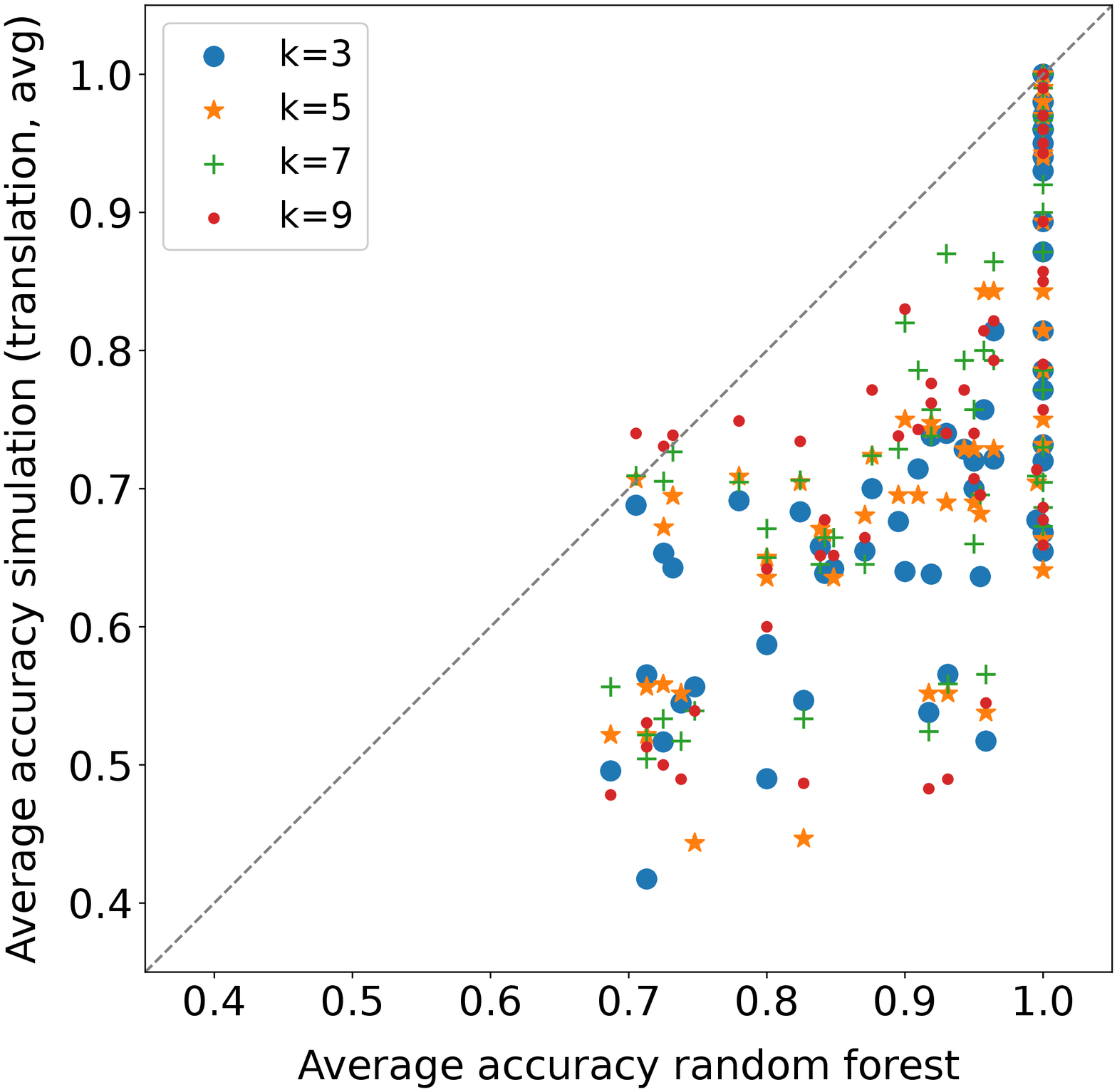}
    } \\
    \subfloat[\label{fig:si-transl-avg-vs-svm-gaussian}]{
        \centering
        \includegraphics[width=0.33\linewidth]{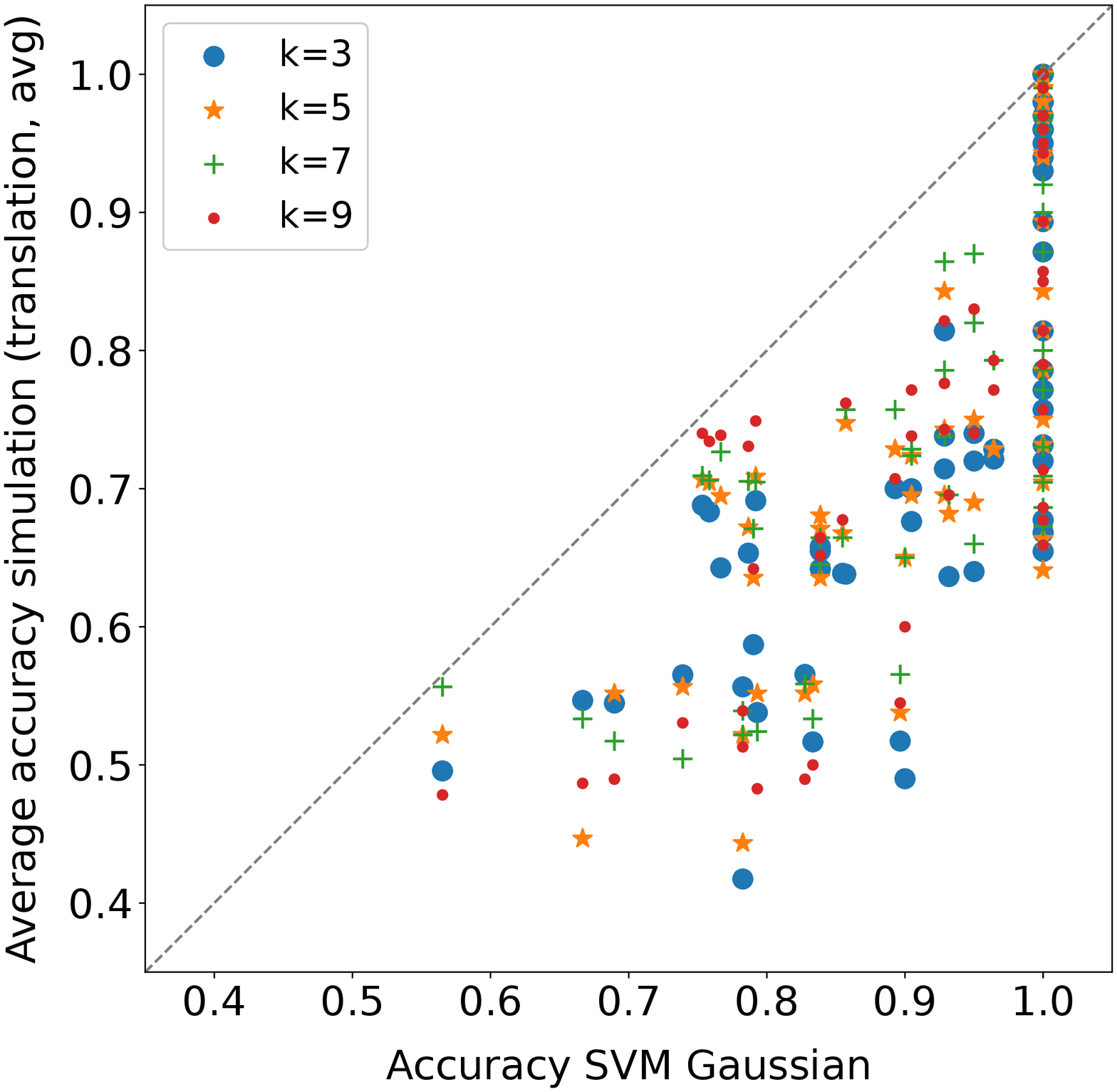}
    }
    \subfloat[\label{fig:si-transl-avg-vs-svm-linear}]{
        \centering
        \includegraphics[width=0.33\linewidth]{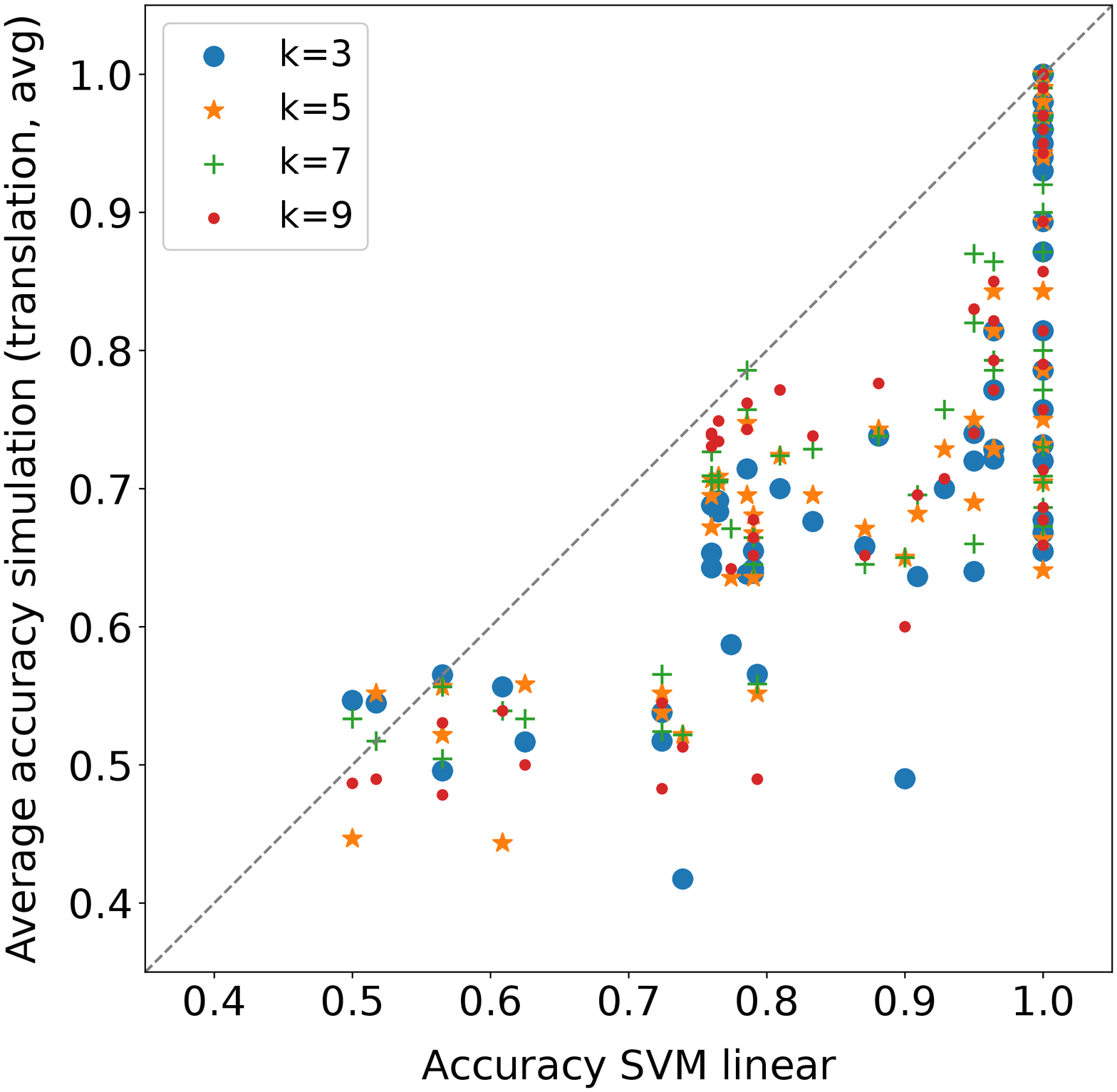}
    }
    \caption{Comparison between some classical baseline methods and \textit{simulation} (\textit{translation}, \textit{avg}) in terms of accuracy. The number of shots for \textit{simulation} is 1024, and each point is related to a dataset fold}
    \label{fig:si-comp-with-baselines}

    \vspace{10pt}

    \hypersetup{hidelinks}
    \centering
    \captionof{table}{Wilcoxon signed-rank test ($\alpha\,{=}\,0.05$) applied to the distributions shown in \cref{fig:si-comp-with-baselines}. The values reported in the table are the p-values obtained}
    \label{tab:si-comp-with-baselines-stats}
    \small
    \begin{tabular}{c|c|c|c|c}
                                                  & \textbf{k=3} & \textbf{k=5} & \textbf{k=7} & \textbf{k=9} \\ \hline
		\cref{fig:si-transl-avg-vs-knn-cos}       & 2.147E-10    & 5.660E-10    & 8.949E-10    & 1.210E-09    \\ \hline
		\cref{fig:si-transl-avg-vs-random-forest} & 1.105E-10    & 2.521E-10    & 3.705E-10    & 6.632E-10    \\ \hline
		\cref{fig:si-transl-avg-vs-svm-gaussian}  & 1.105E-10    & 2.380E-10    & 3.494E-10    & 3.500E-10    \\ \hline
        \cref{fig:si-transl-avg-vs-svm-linear}    & 3.172E-10    & 3.976E-10    & 1.299E-09    & 3.502E-10    \\ 
    \end{tabular}
\end{figure}

\subsection{Number of shots analysis}
\label{subsec:app-num-shots-analysis}

\begin{figure}[h!]
    \centering
    \includegraphics[width=0.50\linewidth]{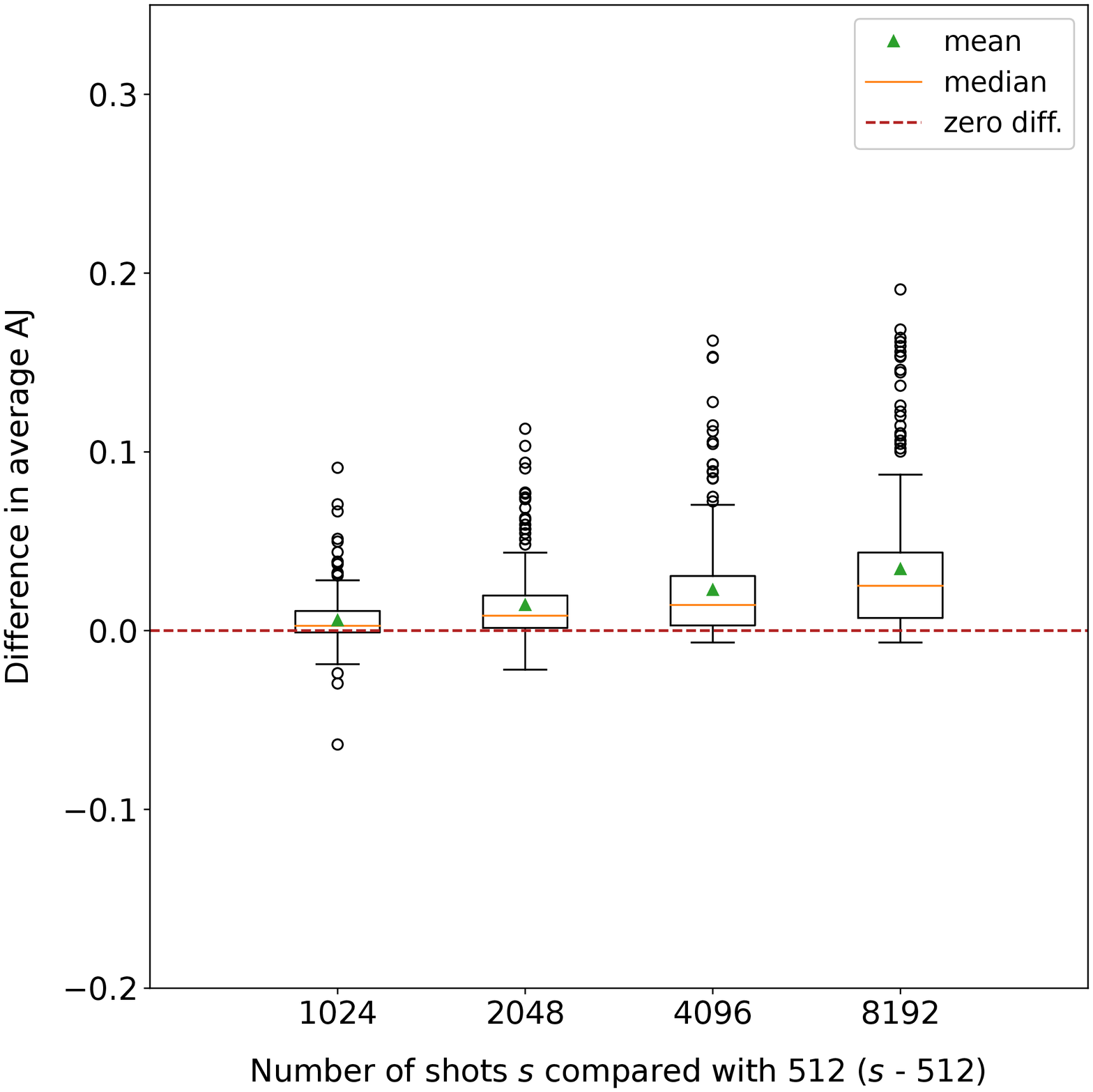}
    \caption{Comparison of different numbers of shots in terms of Average Jaccard score for the \textit{simulation} execution modality in the (\textit{extension}, \textit{avg}) configuration. Each data point corresponds to the difference for a (\textit{dataset fold}, \textit{k value}) pair}
    \label{fig:shots-num-comp-aj}

    \vspace{10pt}

    \hypersetup{hidelinks}
    \centering
    \captionof{table}{Wilcoxon signed-rank test and one-sample T-test applied to the distributions shown in \cref{fig:shots-num-comp-aj}. The values reported in the table are the p-values obtained ($\alpha\,{=}\,0.05$)}
    \label{tab:shots-num-comp-aj-stats}
    \small
    \begin{tabular}{c|c|c|c|c}
                 & \textbf{1024-512} & \textbf{2048-512} & \textbf{4096-512} & \textbf{8192-512} \\ \hline
        Wilcoxon & 1.977E-11         & 8.804E-30         & 2.327E-37         & 1.030E-39         \\ \hline
        T-test   & 3.037E-09         & 6.719E-23         & 1.862E-27         & 3.622E-33         \\
    \end{tabular}
\end{figure}
\vspace{-5pt}

\end{appendices}

\end{document}